\documentstyle[11pt,aaspp4,flushrt]{article}

\def\sun{\hbox{$\odot$}}

\begin{document}

% declarations for front matter

\title{Energy Input and Mass Redistribution by Supernovae in the
Interstellar Medium}

\author{K. Thornton\altaffilmark{1}}
\affil{The University of Chicago}
\authoraddr{The Department of Astronomy and Astrophysics, 
The University of Chicago, 5640 S. Ellis Avenue, Chicago, IL 60637}
\author{M. Gaudlitz\altaffilmark{1}, H.-Th. Janka\altaffilmark{1}, 
and M. Steinmetz\altaffilmark{1,2}}
\affil{Max-Planck-Institut f\"ur Astrophysik}
\altaffiltext{1}{E-mail: kat@oddjob.uchicago.edu, gaudlitz@MPA-Garching.MPG.DE,
thj@MPA-Garching.MPG.DE, msteinmetz@as.arizona.edu.}
\altaffiltext{2}{now at Steward Observatory, 
University of Arizona, Tucson.}
%\begin{document}
% typeset front matter
% \maketitle
\begin{abstract}

We present the results of numerical studies of supernova
remnant evolution and their 
effects on galactic and globular cluster
evolution. 
We show that parameters such as the density
and the metallicity of the environment
significantly
influence the evolution of the remnant,
and thus change its effects on
the global environment (e.g., globular clusters,
galaxies) as a source of thermal and kinetic energy.

We conducted our studies using a one-dimensional
hydrodynamics code, in which
we implemented a metallicity dependent cooling function. 

Global time-dependent quantities such as the
total kinetic and thermal energies and the radial
extent are calculated for a grid of parameter sets. The quantities calculated
are the total energy, the kinetic energy, the thermal energy, the radial
extent, and the mass.  We distinguished between the hot, rarefied
bubble and the cold, dense shell, as those two phases are distinct in
their roles in a gas-stellar system. 

We also present power-law fits to those quantities as a function of
environmental parameters
after the extensive cooling has ceased.  The power-law fits enable
simple incorporation of improved supernova energy input and matter 
redistribution 
(including the effect of
the local conditions) in galactic/globular cluster models.

Our results
for the energetics of supernova remnants in the late stages of their
expansion give total energies ranging from $\approx 9 \times 10^{49}$ to
$\approx 3 \times 10^{50}$ ergs, with a typical case being $\approx$ 10$^{50}$
erg, depending on the surrounding environment. About $8.5 \times 10^{49}$ erg
of this energy can be found in the form of kinetic energy.

Supernovae play an important role in the evolution of the interstellar
medium and galaxies as a whole, providing mechanisms for kinetic energy
input and for phase transitions of the interstellar medium.  
However, we have found
that the total energy input per supernova is about one order of magnitude
smaller than the initial explosion energy.
 
\end{abstract}

\section{INTRODUCTION}
\label{intro}
The role of supernovae (SNe) as sources of matter and energy to the 
interstellar medium (ISM) has been confirmed 
by numerous observations and theoretical studies.
It is believed that supernova explosions
are the source of the hot galactic and halo gas that is seen in 
X-rays.  SNe are by far the major source of the 
heavy elements (Woosley \& Weaver\markcite{Woos1986} 1986), and
the study of metal abundances has proven
to be very useful
in tracing the history of our Galaxy and other galaxies.
SNe are
also the major source of the kinetic energy of interstellar clouds.
Aboott\markcite{Abbo1982} (1982) estimated 
that the supernova energy input is larger by about a factor of five
than the combined input from O, B, A, supergiant, and Wolf-Rayet 
stars, assuming Type I and Type II SN energies of 
$5 \times 10^{50} {\rm ergs}$ and $10^{51} {\rm ergs}$, respectively.
Such energy sources influence 
the subsequent 
star formation in the ISM, which in turn changes the
SN rate and the resulting energy input.  The
interactions between the various physical processes in
the ISM complicate studies of the ISM.   
In particular, since supernovae are the major source of 
energy to the ISM, a proper treatment of supernovae
in the modeling of the dynamical evolution of galaxies or globular clusters
is essential.  

There have been many studies of the interactions of supernova remnants (SNRs)
with the ISM and the late stages of remnant evolution.
Over the last decades, much progress has been made in understanding
the behavior and the characteristics of SNRs.
These studies include analytical and numerical models of
various stages of the remnant evolution.
Chevalier and coworkers considered SNRs in a spherically 
symmetric medium (Chevalier\markcite{Chev1974a} 1974, 
\markcite{1984} 1984) 
and a plane-stratified medium (Chevalier \& Gardner\markcite{Chev1974b} 1974). 
More recently, they extended their work to study the instabilities due to 
radiative cooling (Chevalier \& Blondin\markcite{Chev1995} 1995).  
Those studies tended to focus on the SNR evolution itself in an
attempt to explain
the observations of SNR of various ages.  

Other studies focused on the interactions 
between the SNRs and the ISM. 
Cox and Smith\markcite{Cox1974} (1974) suggested that SN 
explosions could create a hot gas phase in the ISM.  
McKee \& Ostriker\markcite{McKe1977} (1977) proposed
that the interstellar medium consists of three phases: a cold neutral
medium, a warm ionized medium, and
a hot ionized medium.  
Slavin \& Cox followed these studies with detailed predictions
of column densities of highly ionized elements, such as \ion{O}{6},
\ion{Si}{4}, and \ion{C}{4} (Slavin \& Cox\markcite{Slav1992} 1992), 
and the porosity 
factor (Slavin \& Cox\markcite{Slav1993} 1993) of 
the solar neighborhood.  
Cioffi, McKee, \& Bertschinger\markcite{Ciof1988} (1988) 
(hereafter, CMB) studied 
the evolution of a SNR using
a one-dimensional numerical hydrodynamical model, which included the 
effects of cooling
by radiation.  The results were later applied to model the ISM 
in a galactic disk (Cioffi \& Shull\markcite{Ciof1991} 1991).  
However, these studies were carried out 
only for the case
of a typical present-day interstellar environment with solar metallicity
(or, at best, of an environment of relatively comparable properties). 
The studies to date have been mainly 
intended to provide an understanding of the
observations of the present-day SNRs and ISM, and thus the results
were not extended for applications to the modeling of formation 
and early evolution of 
stellar systems.  
An example of the exception is the work by Hellsten \& 
Sommer-Larson\markcite{Hell1995} (1995),
who performed numerical simulations and an analytical study to 
calculate the mass fraction of hot gas in supernova remnants for a range 
of ISM densities and for a few choices of metallicities.  The result,
however, was limited to the fraction of hot gas, and provided
no clear method for its application.
Therefore, a systematic study of the effect of SN explosions
in various environments that exist in the course of galactic 
evolution has not been carried out.  As a result, no
realistic prescription for SN energy dispensation is yet available 
to researchers interested in proper modeling of galaxies.

As more sophisticated models of galaxies and other stellar systems
have been developed,
the need has increased for more accurate data which describe the behavior of 
SNe in various environments.  
Due to the lack of such information, simplifying assumptions have been
made in models which involve supernova
heating and kinetic energy input.  A common method of incorporating
SNe energy input is simply to assume a typical SN explosion energy 
of about $10^{51} {\rm ergs}$ (Leitherer, Robert, \& Drissen\markcite{Leit1992} 
1992, Burkert, Hensler, \& Truran\markcite{Burk1992} 1992).  
Due to the uncertainty 
as to how efficiently the explosion energy is transferred to the ISM, 
some studies introduce
a parameter, the ``efficiency'', which measures how
much of this explosion energy becomes available to the ISM.   
The uncertainty in our knowledge of the magnitude of this efficiency 
is reflected in the wide range of assumed values:
from 1\% for kinetic energy (e.g., Padoan, Jimenez, \& 
Jones\markcite{Pado1997} 1997) 
up to 100\% (e.g., Burkert {\it et~al.}\markcite{Burk1992} 1992; 
Theis, Burkert, \& Hensler\markcite{Thei1992} 1992; Rosen and 
Bregman\markcite{Rose1995} 1995).  In 
some cases, the values of efficiency are determined by 
fits to the observation.
There are also studies in which the supernova explosion energy of
$10^{51} {\rm ergs}$ is put purely into the thermal energy 
(Katz\markcite{Katz1992} 1992, Steinmetz \& M\"uller\markcite{Stei1995}
1995).  They concluded that input in thermal energy is easily radiated 
away, and thus has insignificant effect in their models.  
Navarro \& White\markcite{Nava1993} (1993) found a strong
dependence of the evolution of galaxies on the fraction, $f_v$, 
of energy input that
is provided in the kinetic form.   
Cole {\it et. al.}\markcite{Cole1994} (1994) concluded that $f_v$ of 
about 10\% to 20\% provided a good fit to observational data 
(such as the galaxy luminosity functions, galaxy colors, the 
Tully-Fisher relation, faint galaxy number counts, and the redshift
distribution).
To limit the uncertainties in the quantities and the fractions of 
energies that the supernovae provide,
it is important to determine the input from a basic physical 
approach.

SNRs are also known to produce dense, cold environments,
or clouds, in the shell during their late evolution.  This provides an
important site for star formation.  Although such effects have been
studied in terms of ``enhanced star formation rate'' (Ikeuchi \& 
Habe\markcite{Ikeu1984} 1984),
they have not been examined consistently with respect to either 
the energy input 
or to the nature of the environment.

In this paper, we present our approach to this problem,  as well as some
selected results from the grid of models, which provides 
information about the
global characteristics of SNRs in various environments.
We will present the results of numerical simulations for a range of
conditions relevant to 
the entire period of galactic formation and evolution, including
the halo formation period, as well as globular cluster formation.

This paper is organized as follows.
In \S\ref{numsim}, we discuss our numerical simulations 
in some detail.  In \S\ref{results}, we present the results from our 
calculations, focusing on the physical processes involved.  
In \S\ref{fits}, we present a set of power-law fits to our numerical
results of quantities which characterize the effects of SNe in various
ISM environments.
We then  discuss, in \S\ref{discussion},
some of the assumptions we have made in 
the calculations, and possible implications of our results 
for dynamical and chemical evolution of galaxies and globular clusters.
Our conclusions are presented in \S\ref{conclusions}.

\section{NUMERICAL SIMULATIONS}
\label{numsim}

\subsection{Assumptions and Input Physics} 
\label{methods}
The Lagrangian equations governing the spherically symmetric 
hydrodynamical system are:
$${1 \over \rho} = {4 \pi \over 3}{\partial r^3 \over \partial m},$$
$${{\rm d}r \over {\rm d}t} = v,$$
$${{\rm d}v \over {\rm d}t} = - {1 \over \rho} {{\partial P} \over 
{\partial r}} - {G m(r) \over r^2},$$
$${{{\rm d}\varepsilon} \over {{\rm d}t}} = {(P + \varepsilon) \over {\rho}}
{{\rm d} \rho \over {{\rm d}t}} 
- n_e n_H \Lambda (T),$$
$$P = (\gamma-1)*\varepsilon,$$
$$T = {\mu m_H P \over {k \rho}},$$
where $\varepsilon$ is the energy density,
$n_e$ and $n_H$ are the electron and the hydrogen number densities, 
respectively, $\Lambda(T)$
is the cooling function, and $m_H$ is the mass of hydrogen.  
Other variables have their standard definitions.
These equations are solved using the numerical methods described
by Janka, Zwerger, \& M\"onchmeyer\markcite{Jank1993} (1993). 
The boundaries were closed both at the center and at the outer
end.  Additional zones are added as the shock shell approaches the outer
boundary, so that the closed boundary does not affect the evolution.

In this work, we provide an improved measure of the
thermal and kinetic energy input of supernovae to their environments 
for a grid of initial conditions.  In particular,
we examine the effects of varying
the metallicity and the density in the
environment.
The ranges of initial composition and ambient density are chosen to 
provide adequate coverage for early and late galactic environments.
For the metallicities, mass fractions of metal 
$Z = 0.00$, $0.01$, $0.02 (\equiv Z_{\sun})$, and $0.04$
are chosen.
We consider densities, $\rho_0$, ranging from 0.0133 to 13.3
$m_H/{\rm cm}^3$
(corresponding to $n_H$ of $0.01$ to $10.0 {\rm cm}^{-3}$ with $Z = 0.02$).
Wider ranges in both the metallicity and the density are adopted for 
calculations for the power-law fits (see \S\ref{fits}).

The gas is assumed to be composed of 23\%
helium, $(77-Z \times 100)$\% hydrogen and $(Z \times 100)$\% metals
by mass, and to be monatomic and
nonrelativistic, so that $\gamma = 5/3$.  
Each simulation starts with 1800 grid points
distributed over 100 pc, with 150 grid points in
the innermost region where the supernova explosion
energy and the ejecta mass are located initially. 

The initial configurations are:
\begin{itemize}
\item {Outer region (ISM): The density and the metallicity are assumed
constant at the chosen values, which are varied between different 
calculations.}
\item {Inner region (exploding SNR, inner 1.5 pc):  }

\begin{itemize}
\begin{enumerate}
\item {Ejecta Mass:  3 $M_{\sun}$ are distributed uniformly, 
in addition to the
mass contributed by the ISM in this volume.  
The results of our calculations are not strongly dependent upon the 
assumed mass of the supernova ejecta.}
\item {Thermal Energy:  $6.9\times 10^{49}$ ergs are 
distributed uniformly over the region.}
\item  {Kinetic energy:  $9.31\times 10^{50}$ ergs are distributed such that the
velocity profile is linear (similar to the Sedov solution).}
\end{enumerate}
\end{itemize}
\end{itemize}

A critical piece of input physics for our study is the cooling functions.
We adopt the metallicity dependent 
cooling functions calculated by B\"ohringer and
Hensler\markcite{Boeh1989} (1989), which assume optically
thin gas in thermal equilibrium.
This study includes atomic
lines of the ten most abundant elements (H, He, C, N, O, Ne, Mg, Si, S, and Fe)
in the wavelength range 1.5 \AA~to 2340 \AA.  The actual 
cooling rate is given by
$n_e n_H \Lambda (T)$, where $\Lambda (T)$ is the
cooling function, T
is the gas temperature, and $n_e$ and $n_H$ are the
number densities of electrons
and hydrogen, respectively. 

Below the temperature of $10^4$~K, the 
cooling function depends on the trace ionization.  We estimate
the cooling in this regime from the cooling functions of Dalgarno
and McCray\markcite{Delg1972} (1972), using a 
normalization consistent with our work.  
It should be noted that the cooling functions
we have adopted in the low temperature regime ignore molecular and
neutral atomic processes.
This is a possible weakness in this 
work.  However, in order to make our treatment more realistic, 
a detailed study of radiative processes involving molecules and neutral
atoms, including reliable population information for each species,
must be obtained first. 
We have chosen to make conservative estimates of the cooling to provide
a lower limit to the energy lost to radiation.
\begin{figure}[t]
\epsscale{0.80}
\plotone{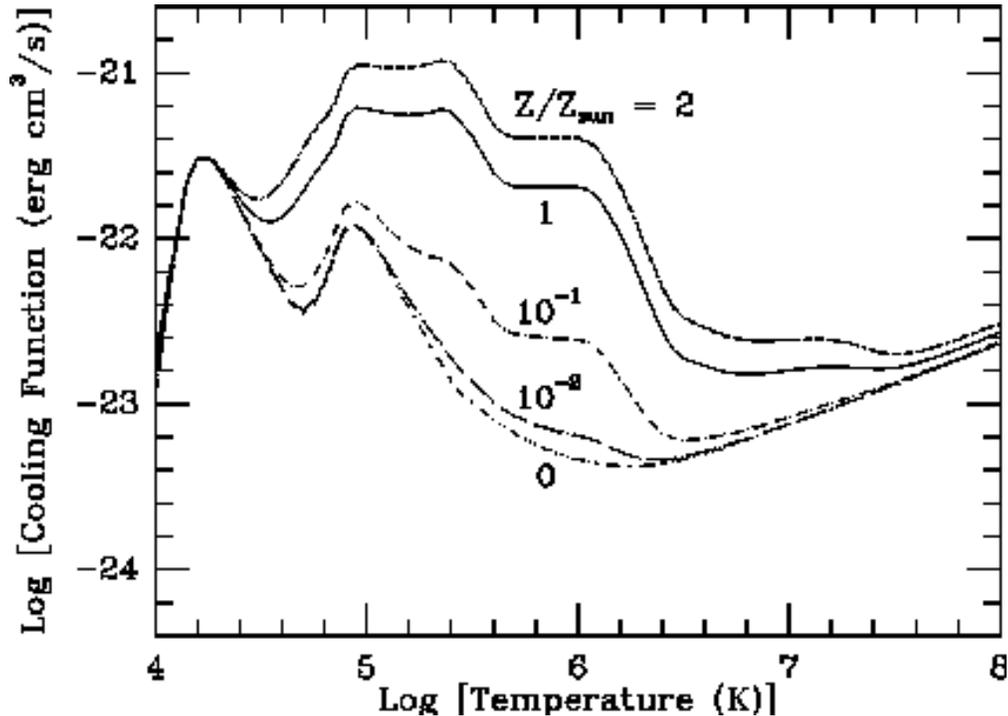}
\caption[f1.eps]
{Cooling functions adopted for the calculation for 
temperatures above $10^4$~K from B\"ohringer \& Hensler (1989).
\label{coolingfunction}}
\end{figure}

The cooling functions we have adopted in our calculations are shown
in Figure \ref{coolingfunction}.  Cooling functions 
simplify the implementation of cooling by collecting the effects
of radiation from many atomic species.   
Only hydrogen and helium contribute to the cooling
for the case of primordial galactic matter, and 
therefore, the coefficient is lower than the case of the solar 
metallicity.  In particular, the metal cooling is very efficient
in the temperature range $10^5$~K to $10^7$~K, and thus even trace
amounts of metals dominate in the temperature range.  For a
solar metallicity environment, the metals dominate the cooling
rate by as much as a
factor of 10 - 100 in the temperature range $10^5 - 10^7$~K.   

It should be noted that the various studies adopted different 
descriptions
for the cooling (both in the cooling coefficient and in its normalization;
a normalization to $n_H^2$, rather than to $n_e n_H$, is
more commonly used).
For example, CMB 
adopted a simple description of cooling
in which the cooling function is proportional to the powers of the metallicity
and the temperature.  This enabled them to solve a simplified
differential equation analytically for the expansion of a cooling 
SNR in an environment with metallicities similar to the solar case.  
However, in the limit of 
very low metallicity, the solution breaks down, since the dominant 
cooling is provided by hydrogen and helium, for which a
simple power-law cooling function does not apply.  
For numerical studies, there are 
various factors which must be considered in calculating
cooling functions, such as the composition of the metal component, 
the radiative transitions to include, and the normalizations (i.e., the
cooling rate is proportional to the product, $n_e n_H$, etc.).  We tested our
results above $T = 10^4$~K using a cooling function
calculated by Sutherland \& Dopita\markcite{Suth1993} (1993), 
which is slightly different from that of
B\"ohringer \& Hensler(1989), and found good agreement. 
For the case with solar metallicity and an ambient density
of $0.1~{\rm hydrogen}/{\rm cm}^3$, 
the comparison between the two results yielded
a difference in the total energy of 2.4\%, or 
4.1$\times 10^{48}$ ergs at the time
the total energy settled to approximately a constant value.  This indicates
that the cooling function above $T = 10^4$~K is known sufficiently well for 
the purposes of this
study.  For temperatures below $10^4$~K, we expect the uncertainty to be larger
since the calculation of the cooling function in the regime is complicated
by molecular cooling. 

The metallicity of the gas, used in calculating the metal-dependent
cooling, is assumed constant at the ambient medium value.  Although
there is an enhancement of heavy elements in the region where mixing
takes place, it has been determined from 2-D hydrodynamics calculations that 
this occurs only in the inner region well away from the shock front
where most of the cooling takes place (Gaudlitz 1996).  
The cooling rate in the very low density bubble, where
possible enhancement of metals occur, is much smaller than that of
the shell, despite the high metallicity.   
Therefore, the small error in 
the local cooling rate in the bubble is negligible.

Following CMB, we assume that the gas is fully ionized.  
The effects of magnetic fields
are ignored in the present calculations; we will consider the possible
consequences of this assumption in our discussion in \S\ref{discussion}.  

It is recognized that the region behind the 
shock is under-ionized,
due to the fact that the ionization time becomes longer than the
local dynamical time.  In order to properly treat this effect, 
time-dependent ionization and recombination must be 
implemented, instead of simply assuming full ionization.
We do not expect that these small modifications to the pressure 
and to the cooling history will change the 
global properties of the SNR significantly.  
It should be kept in
mind that a precise treatment of ionization is essential if a model is
to be used to predict emission spectra from SNRs, which are
sensitive to level populations.  

The effects of thermal conduction may be important 
in the late stages of the SNR evolution.  We expect that conduction
should indeed modify the temperature profile in the SNR.  
However, we do not
expect a significant change in the cooling itself, since the temperature
in the shell (where most cooling takes place) is not affected 
significantly.  Also, it is difficult to quantify the effects of conduction, 
because turbulent magnetic fields are known to suppress 
conduction.  Since we do not have any information on the magnitude
of turbulence or on the strength of the magnetic fields, we have chosen not to 
include the effects of conduction in this study.  

In addition, we 
ignore the kinetic energy loss due to cosmic-ray radiation.   This
energy loss is expected to occur at a very early stage of the SNR evolution; 
therefore, the effect can be taken into account simply by scaling
the results with the initial energy. 

\subsection{Hydrodynamic Code and Test Calculations}
\label{code}
We mainly employed  
an explicit Lagrangian finite-difference 
scheme, which is 
second-order accurate in space and in time, in our numerical studies.
For treating shock discontinuities, a tensor form of the
artificial viscosity (Tscharnuter \& Winkler\markcite{Tsch1979}
1979) is used.
The code has
been tested extensively through standard problems with known solutions, 
and its performance compared well to a PPM code (Janka,
Zwerger, \& M\"onchmeyer\markcite{Jank1993} 1993).  

The energy loss due to radiation
was treated as a source term, which is implemented 
by an operator-splitting method. 
The electron abundance was calculated as discussed previously, and
was used 
both in the equation of state and in calculating the cooling rate.

The time steps are limited so that no quantities change by more than
10\% within a time step, as well as by the time step constraints arising 
from the Courant-Friedrichs-Lewy stability condition, the dynamical 
time, and the cooling time.
\begin{figure}[t]
\epsscale{0.80}
\plotone{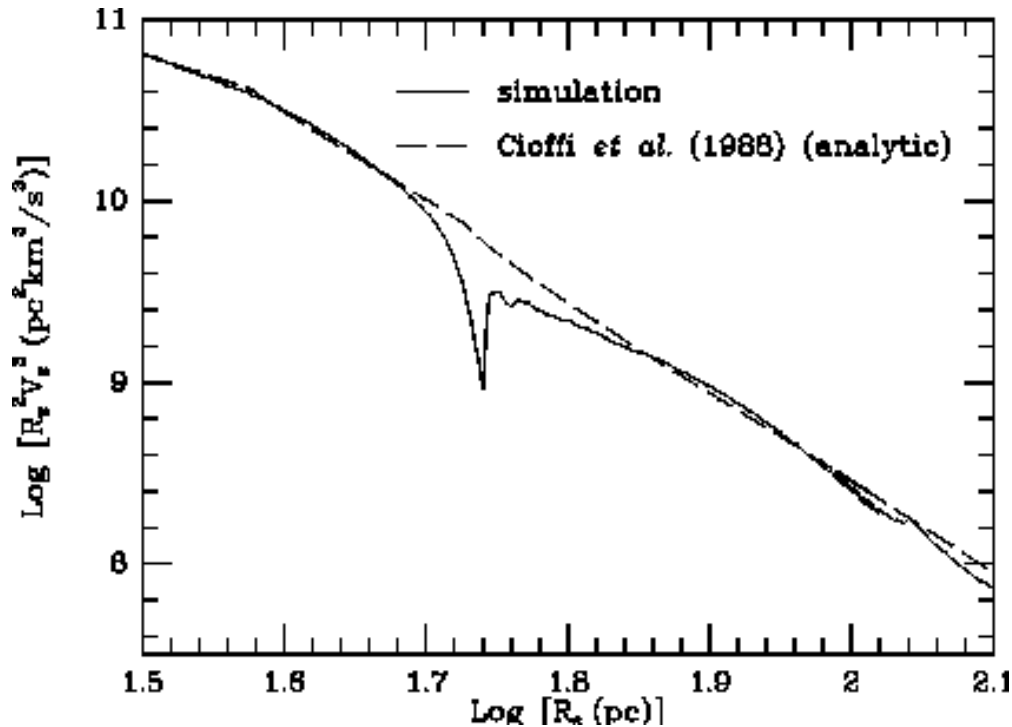}
\caption[f2.eps]
{The product, $R_s^2V_s^3$, is plotted against $R_s$ for
the parameter set $n_H = 0.1 /{\rm cm}^3$ 
and $Z = 0.02$ (solar metallicity).
$R_s$ and $V_s$ are the radius and velocity of the shock. \label{R2V3}}
\end{figure}

As a first test of our calculation, our results are compared to that of 
CMB for an appropriate
parameter set ($n_H = 0.1 /{\rm cm}^3$, $Z = 0.02$).  
We found the results to be consistent within the 
expected difference due to the choice of the cooling functions.  
An example is given in Figure \ref{R2V3}, which
should be compared with  
Figure 8 of CMB. 
The plotted quantity is $R_s^2V_s^3$ 
vs.~$R_s$, where $R_s$ is the radius at which the shock
is located, and $V_s$ is the shock velocity.  
The quantity $R_s^2V_s^3$ is closely related to
the luminosity, as it is equal to the decrease in kinetic energy,
which is approximately ${1 \over 2} \dot{M} V_s^2 $.
The second term in the expression
for the rate of kinetic energy change 
becomes negligible as the time increases, since it is
proportional to $\dot{V}_s$, which is a rapidly decreasing function of time.
The dotted lines are the analytic solution
given in CMB.
The analytic solutions in the two plots differ slightly.  We have used
the following expressions for the radius $R_s$ and velocity $V_s$ of the
shell, as provided by CMB:
$$R_s = R_{PDS} {\left( {4 \over 3} t_* - {1 \over 3}
\right)}^{3/10},$$
$$V_s = V_{PDS} {\left( {4 \over 3} t_* - {1 \over 3}
\right)}^{-7/10},$$
where 
$$R_{PDS} = 14.0 {{E_{51}^{2/7}} \over {n_H^{3/7} \zeta^{1/7}}} pc
% = 37.6 pc,$$
,$$
$$V_{PDS} = 413 n_H^{1/7}{E_{51}^{1/14}}\zeta^{3/14} km/s 
% = 297 km/s,$$
,$$
$$t_* = {t \over t_{PDS}},$$
$$t_{PDS} = 1.33 \times 10^4 {{E_{51}^{3/14}} \over 
{\zeta^{5/14} n_H^{4/7}}} yr.$$

The subscript PDS indicates the quantities at the onset of the 
pressure-driven snowplow phase, i.e., at $t = t_{PDS}$.
$n_H$ is the ambient hydrogen density which was set to $0.1/{\rm cm}^3$ 
in both
calculations.  $\zeta$ is the metallicity parameter, $Z/Z_{\sun}$, which 
was set to 1.  $E_{51}$ is the initial explosion energy in $10^{51}$ ergs.
CMB used a slightly different value of 
$R_{PDS}$ (36.8 pc, as opposed to 37.6 pc as given by the equation for 
$n_H = 0.1/{\rm cm}^3$), 
corresponding possibly to a slightly different value of $n_H$, 
$0.105/{\rm cm}^3$.
It is clear that our results closely follow 
their analytic solution and numerical
solution.  The slight differences are most likely due to the differences
in the adopted cooling functions.

The extreme thinness of the shell at the time the cooling is most efficient 
can cause numerical problems.  Special care was given to the region
at and around the thin shell to make sure that
our calculation were well resolved at all times. 
This was achieved by a combination of visual inspections of the shell 
region
and by running test cases with higher and lower resolutions.

\section{RESULTS OF SNR EVOLUTION}
\label{results}
The numerical results of our calculations of supernova remnant evolution are 
presented in this section.
We calculate the total
energies, the kinetic energies, and the thermal energies of the SNR models,
differentiating shell energies from bubble energies.  
We also calculate the radial extent of SNR, and the SNR and 
the shell masses as functions of time.  These quantities provide 
the information necessary for proper dispensation of energies and matter
in stellar system formation/evolution models.  The boundaries of the 
shell were determined by over-density of 10\%, as compared to the unshocked
medium.  Test calculations were performed with another selection criterion, 
which separates the hot bubble and the cold shell at the temperature of 
$10^5$~K, and we found no significant differences. 

We are interested in metallicities ranging down to the low values 
characteristic of halo environments (see e.g. the reviews by Wheeler, Sneden,
\& Truran\markcite{Whee1989} (1989)). 
Observations have identified stars 
with metallicities as low as $[{\rm Fe}/{\rm H}] = -3$ to $-4$ (McWilliam, Preston, Sneden,
\& Searle\markcite{McWi1995} 1995,
Ryan, Norris, \& Beers\markcite{Ryan1996} 1996).  
In addition, observations of QSO absorption systems indicate metallicities
as low as $[{\rm Fe}/{\rm H}] = -2$ to $-2.5$, 
which may correspond to the metallicities
of protogalactic environments (Cowie, Songaila, Kim, \& 
Hu\markcite{Cowi1995} 1995; Rauch, Haehnelt, \& 
Steinmetz\markcite{Rauc1997} 1997; Timmes, Lauroesch, \& Truran 1995;
Lauroesch, Truran, Welty, \& York\markcite{Timm1996} 1996; 
Lu, Sargent, Barlow, Churchill, \& Vogt\markcite{Lu1996} 1996; 
Pettini, King, Smith, \& Hunstead\markcite{Pett1997} 1997).
We have
therefore included $\log[Z/Z_{\sun}]$ down to $-3$.  The results from
the lowest metallicity case is applicable also for a zero metallicity 
environment, as explained later.  The density ranges are taken from
the cold cloud like conditions of $n_H = 10^3 /{\rm cm}^{3}$ 
to a very hot rarefied 
gas of $n_H = 10^{-3} /{\rm cm}^{3}$.  
These metallicity and density ranges 
should suffice for the application of our results to diverse star-forming 
environments.

\subsection{Behavior of Global Quantities and 
Details of Remnant Structure and Evolution}
\label{generalresults}
We will first provide an overview of the behavior of the global quantities
such as the total energy and the radius of the SNR in various phases of 
SNR evolution. 
We will then present the structural information
which illustrates the physical mechanisms governing the behavior
of the SNR in those phases.  For this purpose, we will take
representative snapshots from the phases of the remnant
evolution typical of the present-day interstellar environment:
$Z = 0.02$, $n_H = 0.1 /{\rm cm}^3$.

The ejecta-dominated phase, a very early phase in SNR evolution where
the ejecta mass dominates the swept-up ambient matter, is not
studied here since the calculations do not properly simulate it.
During this phase, the SN ejecta expand into 
space much like expansion in a vacuum, since the surrounding matter
does not influence the system significantly.   
Our calculation does not provide
realistic information on this phase,
since the results at the time
corresponding to the end of the phase
are still influenced by the initial conditions, and in some high
density cases,
the calculations are started at conditions corresponding to those
occurring after this
phase has effectively ended.  
Since our focus is on the effect of SNRs on the surrounding ISM, the 
details of the ejecta-dominated phase
(when the influence of the SNR is still confined to a small region of
the ISM)
are not of direct interest for this paper.  
The phase has been well studied in order to
explain X-ray emissions from young SNe, and we 
refer the readers to previous studies (Hamilton \& 
Sarazin\markcite{Hami1984} 1984;
McKee \& Truelove\markcite{McKe1995} 
1995 (this paper also contains discussion of the Sedov-Taylor
phase and of the transition between the two);
Spicer, Clark, \& Maran\markcite{Spic1990} 1990; Kazhdan \& 
Murzina\markcite{Kazh1992} 1992).
Our results properly represent the SNRs starting
at the adiabatic expansion phase.  We will now describe our results
and the physical processes which dictate the observed behaviors.  
\begin{figure}
\epsscale{0.80}
\plotone{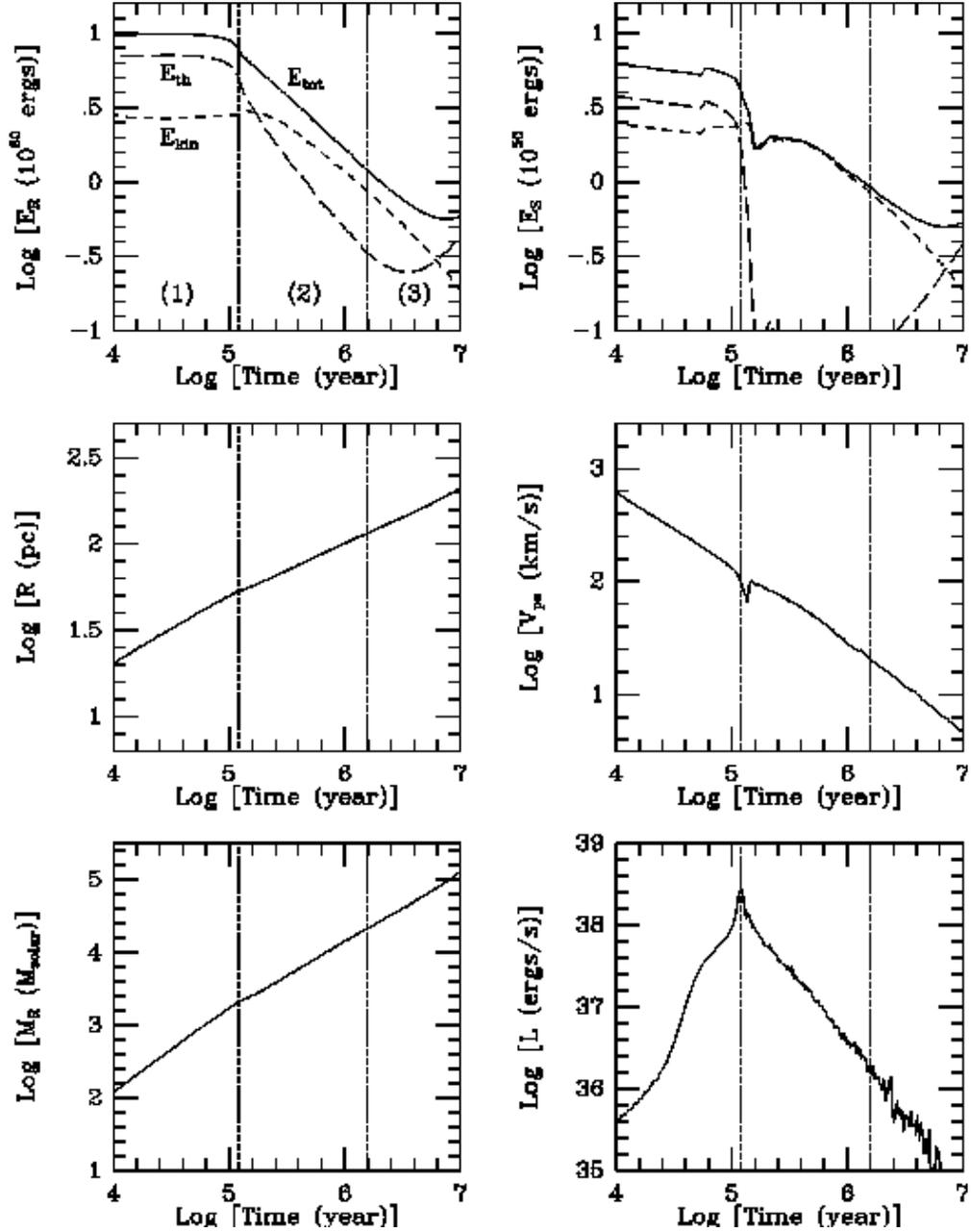}
\caption[f3.eps]{Total energy, kinetic energy,
and thermal energy of the SNR ($E_{Rtot}$ (solid line), $E_{Rkin}$ 
(dashed line),
and $E_{Rth}$ (long dashed line), respectively) and
of the shell ($E_{Stot}$ (solid), $E_{Skin}$ (dash), and $E_{Sth}$
(long dash), respectively) vs.~time,
illustrating the evolution of the SNR.  The
ambient density is taken to be $\rho_0 = 0.133 m_H/{\rm cm}^3$, and the
metallicity is set to $Z_{\sun}$.
In addition, the radius $R$
and the mass $M_{R}$
of the SNR, post-shock fluid velocity $V_{ps}$, and the 
luminosity $L$ (or the energy
loss rate) of the SNR are plotted.  The vertical dotted lines indicate the
approximate phase boundaries (see text). \label{timeevolstd}}
\end{figure}

We will divide the evolution of a SNR (subsequent to 
the ejecta-dominated phase) into three phases, according to
the governing physical processes: 1) the adiabatic (or Sedov-Taylor)
phase (Sedov\markcite{Sedo1959} 1959; Taylor\markcite{Tayl1950} 1950),
 2) the cooling (radiative shock, or pressure-driven snowplow) 
phase (Cox\markcite{Cox1972} 1972; Chevalier\markcite{Chev1974a} 1974), 
and 3) the post-cooling phase.  Evolution of SNRs beyond the
post-cooling phase is not a focus of this paper; therefore, we will
only mention the possible fate of SNR in \S\ref{late} briefly.
These evolutionary phases are roughly indicated by the numbers 
in Figure \ref{timeevolstd}, 
which shows various global quantities as functions of time.
The total energy, the kinetic energy,
and the thermal energy of the SNR and of the shell, 
the radius $R$
and mass $M_{R}$
of the SNR, the post-shock fluid velocity, and the luminosity (or the energy
loss rate) $L$ of the SNR, are shown.  
The boundaries between the phases are noted on each plot.

The curves describing the changes in the energetics of the SNR
allow us to distinguish an early phase
where cooling does not affect the structure (i.e.~cooling
is still negligible and the shell is
still thick), and a phase where cooling has become important and a thin shell
has formed.  This can be seen in the behavior of the total energy plotted
in Figure \ref{timeevolstd}, as a flat plateau at early
times, followed by a rapid energy decrease.
The third phase, the post-cooling phase, is seen as a flattening of the 
total energy curve after the rapid decrease.  These phases are 
briefly examined below, along with the structural information.

\begin{figure}
\plotone{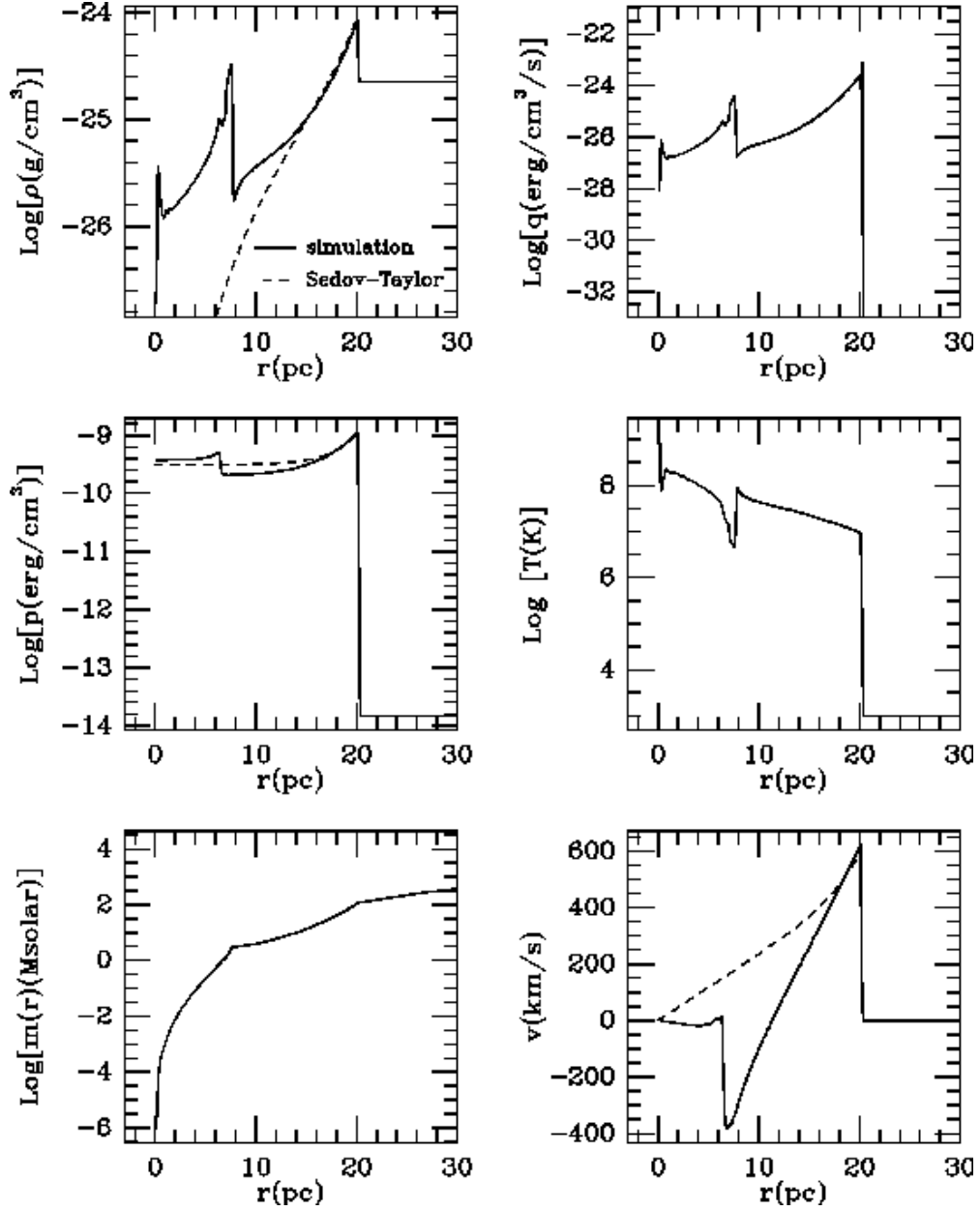}
\caption[f4.eps]{Density $\rho$, pressure $p$, cumulative mass $m(r)$,
cooling rate per unit volume $q$, temperature $T$, and
fluid velocity $V$ 
as a function of radius $r$, at time $t = 9810$ yr. The dashed lines 
represent Sedov-Taylor solution. The
ambient density is taken to be $\rho_0 = 0.133 m_H/{\rm cm}^3$, and the
metallicity is set to $Z_{\sun}$.
\label{str9810}}
\end{figure}

\begin{figure}
\epsscale{0.80}
\plotone{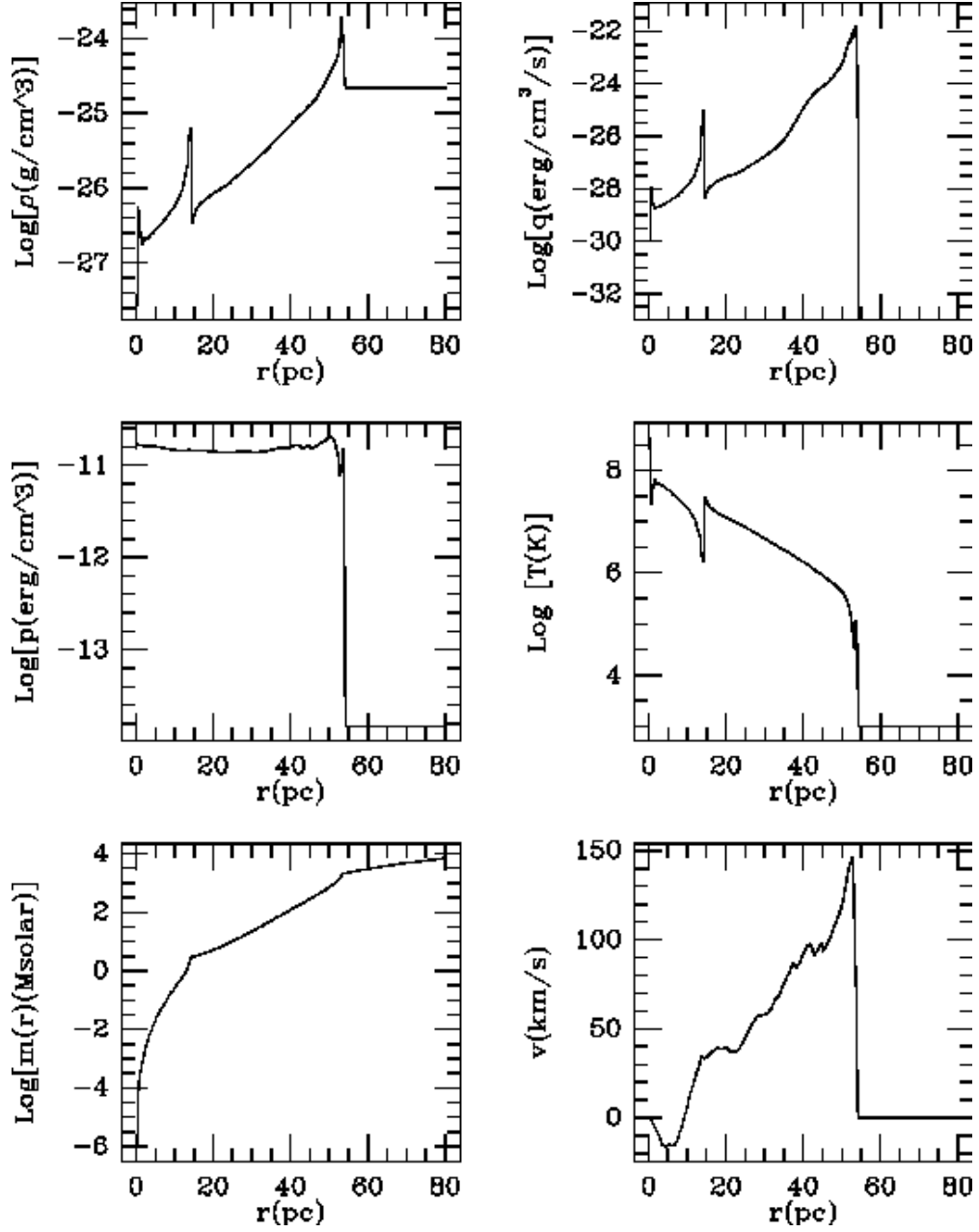}
\caption[f5.eps]{
Density $\rho$, pressure $p$, cumulative mass $m(r)$,
cooling rate per unit volume $q$, temperature $T$, and
fluid velocity $V$
as a function of radius $r$, at time $t = 1.27 \times 10^5$ yr.
The
ambient density is taken to be $\rho_0 = 0.133 m_H/{\rm cm}^3$, and the
metallicity is set to $Z_{\sun}$.
\label{str1.27e5}}
\end{figure}

\begin{figure}
\epsscale{0.80}
\plotone{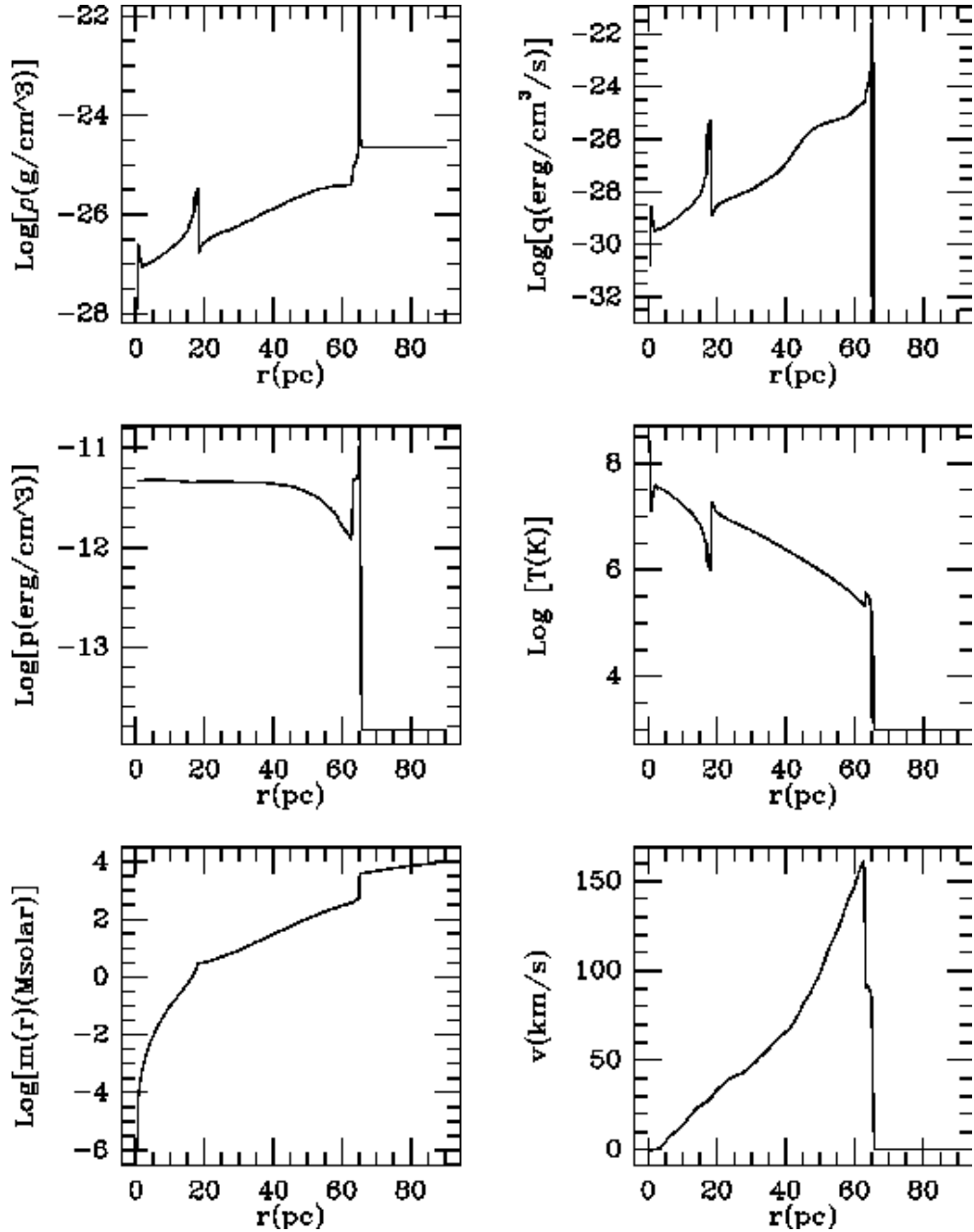}
\caption[f6.eps]{Same as Fig.~\ref{str1.27e5} at $t = 2.54 \times 10^5$ yr.
\label{str2.54e5}}
\end{figure}

\begin{figure}
\epsscale{0.80}
\plotone{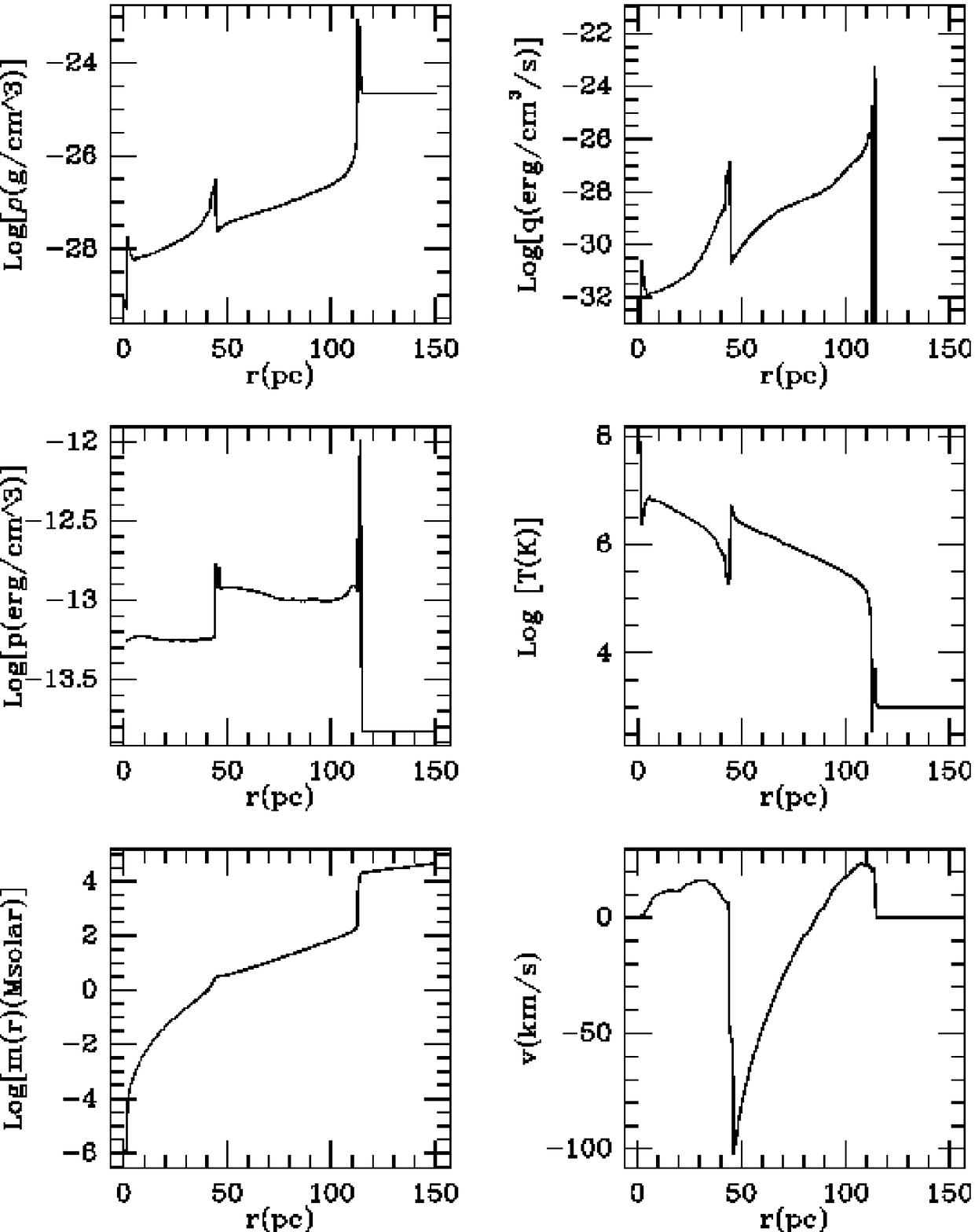}
\caption[f7.eps]{Same as Fig.~\ref{str1.27e5} at $t = 1.52 \times 10^6$ yr.
\label{str1.52e6}}
\end{figure}

The representative structural information is given in Figures 
\ref{str9810} though \ref{str1.52e6}.
Quantities such as density $\rho$, pressure $p$, cumulative mass $m(r)$,
luminosity
per unit volume or cooling rate $q$, temperature $T$, and  
fluid velocity $V$, are plotted 
as a function of radius $r$, at times $t = 9810$, $1.27 \times 10^5$, 
$2.54 \times
10^5$, and $1.52 \times 10^6$ yr.  These times were chosen to represent 
the various phases in the evolution of the SNR.  In addition, the 
Sedov-Taylor solution is plotted with the numerical results of 
density, 
pressure, and post-shock fluid velocity for $t = 
9810$ yr, in order to illustrate the agreement and the disagreement 
between our 
numerical results and the Sedov-Taylor solution.

Although some of these phases have previously been studied, we will  
briefly review the characteristics 
and governing physical processes in each phase.  

\subsubsection{Sedov-Taylor Stage}
\label{ST}
An early phase, where cooling has not been efficient, is represented in 
Figure \ref{str9810} (the structure at $t = 9810$ yr) and
Figure \ref{str1.27e5} (at $t = 1.27 \times 10^5$ yr).  
The structure is similar to that of the Sedov-Taylor solution
(Sedov\markcite{Sedo1959} 1959, Taylor\markcite{Tayl1950} 1950),
which is applicable to the 
adiabatic expansion of a spherical wave (i.e. explosion starting at
an infinitesimally small radius).  
However, a slight indication of the effect of
cooling is already
seen at the shock front by $t = 1.27 \times 10^5$ yr; 
the velocity, density, and pressure are
smaller than predicted by the Sedov-Taylor 
solution.  To demonstrate that this deviation
is due to cooling, the velocity, the density, and the pressure 
profiles at an earlier time, when practically no cooling has yet occurred, 
are compared to the Sedov-Taylor solution in Figure \ref{str9810}.  
It shows the numerical results (solid lines) and the Sedov-Taylor
solution (taken at $t = 9808 + 170$, the approximate age of the SNR 
at the initial condition, dashed lines).  
The peak values 
and the location of the peak values agree well, even though 
the detailed structure deviates because the numerical solution is influenced by
the initial condition of the explosion (with a finite radius),
as expected for a realistic calculation.  This can be seen 
easily in the velocity profile, which shows the
reverse shock propagating inward.  The reverse shock travels
to the contact discontinuity, where it is partially transmitted and reflected,
and to the center, where it is reflected.  The resulting waves
eventually catch up and interact with the shock front, influencing
the shock structure and the cooling history. 
Dynamic relaxation can be achieved
if much of the initial explosion energy is provided 
in the form of the thermal energy in a very small volume
(thus increasing the sound velocity and shortening the relaxation time), as
seen in the results of Chevalier (1974). 
However, previous studies have shown that the bulk of the explosion energy
is put into the motion of the matter, rather than into the 
thermal energy (see CMB and the references
thereof).  Therefore, for realistic initial conditions, 
such as those considered
in our study, we do not expect a complete agreement between the numerical 
results and 
the Sedov-Taylor solution.

As 
mentioned earlier, the Sedov-Taylor solution for adiabatic expansion describe 
the global behavior of quantities like those shown in  Fig.~\ref{timeevolstd}
quite well.  The slope of $\log(R)$ as 
a function of $\log(t)$ from the numerical 
calculation is indeed about 2/5 (0.400 with residual sum $= 3.45 \times
10^{-4}$ for the fit with data between $\log(t[{\rm yr}]) = 3.7$ and $4.9$
), as predicted from the analytical solution.
The post-shock fluid velocity is related to the shock velocity 
$V_{s}$ by
$${V_{ps} \over V_{s}} = {{\gamma - 1} \over {\gamma + 1}}
+ {2 \over {(\gamma + 1}) M^2},$$
where $\gamma$ is the effective adiabatic exponent of the gas, and 
$M$ is the Mach number of the shock, $M^2 = V_s^2/c_s^2$.  
For a strong shock, the second term is negligible.
If there
is negligible cooling, the value of $\gamma$ is the same as the ratio of
specific heats at constant pressure and at constant volume, $C_p/C_V$,
assumed to be 5/3.
This is true during the adiabatic phase, and the ratio 
$V_{ps}/V_{s}$ stays approximately constant; therefore,
the slope of $\log V_{ps}$ is equal to $\log V_{s}$.  The
slope of $\log V_{ps}$ as a function of $\log t$ in Fig.~\ref{timeevolstd} is
about 3/5,  consistent with the prediction for $\log V_{s}$ by the
analytical solution.

The energies stay approximately constant over this phase, with only slight 
adjustment in kinetic and thermal energies once they reach the
equilibrium value.  The distinction between the 
shell and the bubble is not rigorous during this phase, as the thin shell
has not yet formed.  Therefore, the values associated with the shell or
the bubble should be taken with caution.  It is clear from the plots
that the dynamical behavior of the SNR is not influenced until the cooling is 
near the maximum.  

\subsubsection{Radiative Phase}
\label{radiative}
Toward the end of the Sedov-Taylor phase, the effect of cooling 
in the density-enhanced shock front region gradually 
becomes significant and begins to influence the dynamical evolution.  
The pressure just behind
the shock front decreases due to the temperature drop.  The system reacts
by adjusting the velocity profile.  The decrease of the velocity at
the shock front compared to the peak velocity value, as seen in the 
structure at the onset of the radiative phase ($t = 2.54 \times 10^5$ 
yr, Figure \ref{str2.54e5}), is 
due to this effect.  The temperature drop due to cooling is also clearly
seen in the same plot.  
Due to the pressure drop behind the shock front
caused by cooling,
the velocity at the shock front decreases.  The deceleration is not
as large away from the shock front, and therefore the velocity
near the shock front creates a tier, where a reverse shock forms
(see Fig.~\ref{str2.54e5}).
The reverse shock, which appears as a cusp in the velocity profile
just behind the shock front,
travels inwards
relative to the shock front
to the contact discontinuity (see 
Fig.~\ref{str1.52e6}),
reflecting and transmitting at the discontinuity.  The transmitted 
wave travels to the center, where it is reflected.  The wave reflected
at the contact discontinuity travels back toward the shock front.
These waves eventually interact with 
others to create  
complex wave patterns over the entire SNR in its late evolution.

As the cooling becomes very efficient and the thin shell forms at the shock
front, the remnant moves into the radiative phase.  The thermal energy,
converted from kinetic energy is radiated away immediately.  The density 
enhancement at the shock front becomes significantly more than the 
adiabatic (strong shock) value of $(\gamma + 1) / (\gamma - 1)$.
This, in turn, enhances
the cooling, which is proportional to the square of the local density.
This brings the catastrophic cooling.  Much SNR energy is
lost in this phase.  

It should be emphasized that the fraction of the energy 
input from a SN to the ISM,
which is
retained by the ISM, in a solar-like environment, is significantly less
than the explosion energy of the SN.  It is important to realize that
most of such violent energy input escapes in radiation, and therefore
does not provide as much energy input to the ISM as it is often assumed.
Any study which must include such energy input to the ISM in order to model
an evolving stellar system 
must take into account the radiation
loss from the shells of SNRs.
We will consider this point in more detail in \S\ref{discussion}.

\subsubsection{Late Phases: Post-Cooling and Momentum-Conserving Snowplow}
\label{late}
In the very late stage of the SNR remnant evolution, the shell (where
most of the cooling takes place) becomes cold and less dense (due to the
weakness of the shock and the cooling), and consequently 
the total cooling in the SNR becomes 
less efficient.  This phase was not studied by 
CMB since for their analytical study they 
assumed a simple power-law cooling
function which increases as the temperature decreases (and thus
the gas is cooled efficiently to zero temperature, in effect).  Also,
they only followed the remnant evolution to $1.75 \times 10^6$ yr.
Although the effects of shell cooling have become small by this time,
the resulting change in the behavior of SNR characteristics 
is not obvious until about $4 \times 10^6$ yr,
at which time the accreted thermal energy from 
the ambient matter becomes larger than the remnant cooling.
Only then, the resulting increase in the total thermal energy is clearly seen 
(See Fig.~\ref{timeevolstd}).

In the post-cooling phase, the cooling still 
continues in the bubble; however, it does
not become very efficient due to the low density of the gas outside the 
shell.  The total energy is again (approximately) conserved,
as it was in the adiabatic phase. 
The shell becomes thicker as the rate of cooling in the shell decreases.  
Eventually, the 
cooling rate becomes orders of magnitude less than the peak value, and
therefore it can be deemed negligible.  A representative structure at the
transition into this phase 
($ t = 1.52 \times 10^6$ yr) is shown in Figure \ref{str1.52e6}.  
This is very close
to the time at which the final results were taken 
for the fits presented in the later section.  
The structure is characterized by a thick shell with a size of a few 
parsecs and a 
complex velocity profile due to wave interactions.  The bubble is still
hot (T $\approx$ $10^6$~K) and very rarefied ($\rho \approx 10^{-28}$
to $10^{-27} {\rm g}/{\rm cm}^3$).  

The pressure inside the SNR still exceeds the unshocked pressure, although
the difference decreases with time.  The time at which the interior pressure
is no longer significantly larger than the unshocked ambient pressure
depends on the temperature, as well as on the density of its environment.  
At this time, the SNR moves into the momentum-conserving snowplow phase.

In the momentum-conserving snowplow phase, unlike 
the pressure-driven snowplow (or radiative) phase discussed earlier,
there is no longer 
a ``push'' from the interior pressure, since the interior
and exterior pressures are approximately equilibrated.  
The momentum is conserved as the SNR continues to evolve.
The increases
in the total energy and thermal energy at very late
times, seen in Fig.~\ref{timeevolstd}, are due to the accumulation of the
matter and the associated
thermal energy in the ambient medium, which is no longer cooled
by radiation as it becomes part of the SNR.
An analytic solution 
for this phase is easily obtainable from the equations of motion,
energy conservation, momentum conservation, and the equation of state.
Our results are not influenced by the existence of this phase because
the final results (i.e., when most of the cooling is finished) are taken
well before the SNR enters this phase, for a typical ISM pressure.

In some cases, a SNR may become indistinguishable with 
the ISM (i.e., merge with the ISM) 
before it reaches the momentum-conserving 
snowplow phase.  In any case, all SNR will merge with the ISM eventually.
The most common criterion used to determine when 
the transition occurs is the equality of the shock velocity with the
sound velocity of the ISM.  The transition time is thus
dependent on the temperature
of the ambient medium, as well as on the density.  

Our careful examination of the SNR characteristics over the lifetime,
combined with the knowledge of treatments of energy input in stellar system 
formation and evolution, enabled us to determine the best time to 
take the final characteristics of SNRs.  In essence, we have chosen the earliest
time at which the enhanced shell cooling due to radiation
has ceased in effect: 
sufficiently early
so that late time effects such as the accumulation of ambient 
thermal energy are small, and sufficiently late so that the luminosity
has dropped to a small value.  We also used the fact 
that many SNR characteristics
scale well with $t_0$, the time at which the maximum luminosity is attained.
We have determined from the calculations that all models have luminosities
that are less than 0.5\% of the corresponding maximum luminosities 
by the age of $13 \times t_0$.  Therefore, we have chosen to define
the final age, $t_f$, to be $13 \times t_0$.  
At $t_f$, the amount of thermal
energy which has
accumulated from the surroundings (which behaves like the accumulation
of the mass) is well below 5\% of the thermal energy, and below 1\% of
the total energy.  The final time $t_f$,
as defined, occurs earlier than both the onset of the momentum-conserving
snowplow phase and the merging of the SNR with the ISM, for most
interstellar conditions. 
 
\subsection{The Effect of Ambient Density}

\label{densityeffect}
\begin{figure}
\epsscale{0.80}
\plotone{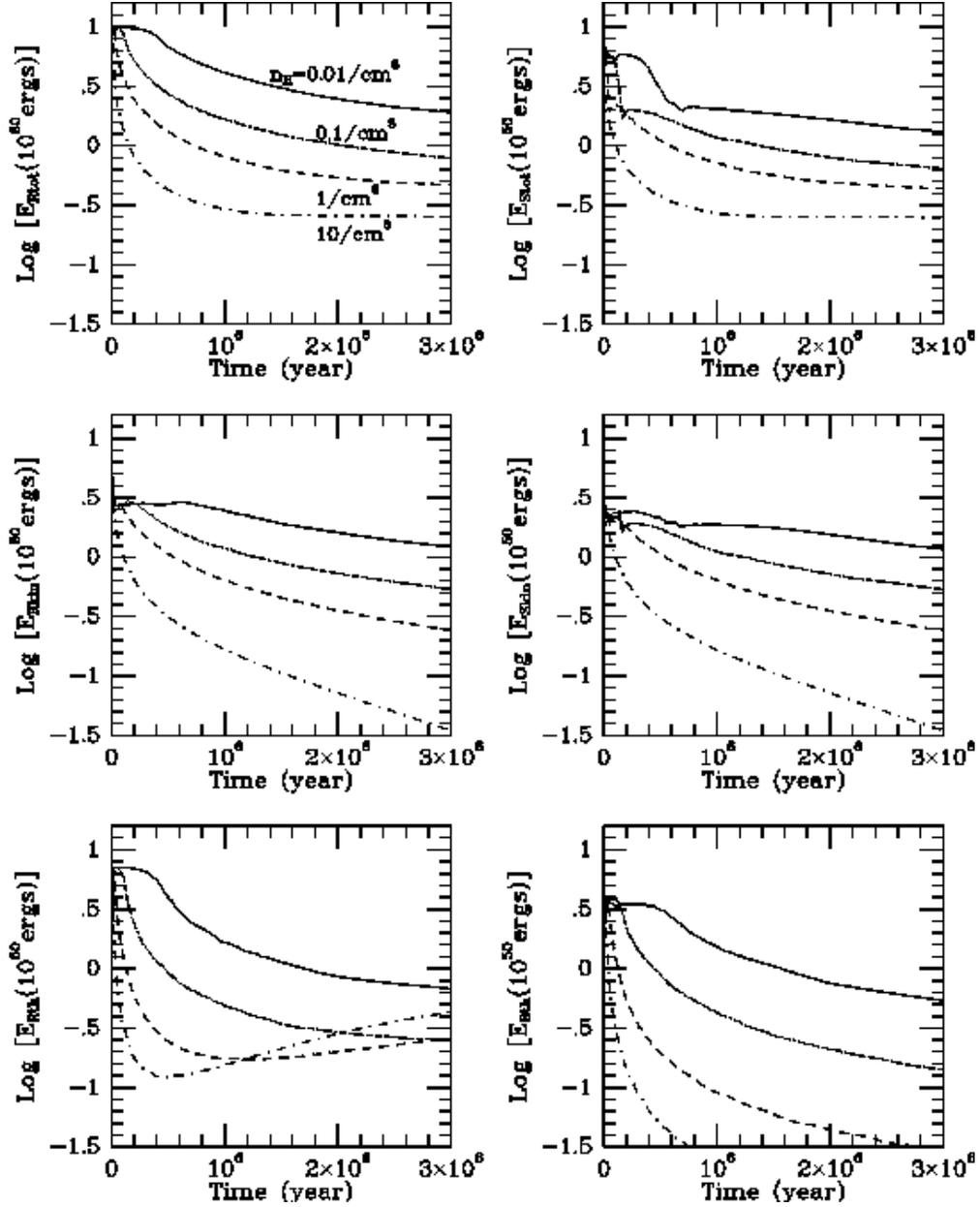}
\caption[f8.eps]{Total energy, kinetic energy,
and thermal energy of the SNR ($E_{Rtot}$, $E_{Rkin}$,
and $E_{Rth}$, respectively) and
total energy, and kinetic energy of the shell ($E_{Stot}$ and
$E_{Skin}$,
respectively), and thermal energy in the hot
bubble $E_{Bth}$ vs. time,
for various cases of ambient density 
($n_H = 0.01$ (solid line), $0.1$ (dotted line), $1.0$ (dashed line), and 
$10.0$ (dash-dotted line) ${\rm cm}^{-3}$).  The metallicity
is fixed at $Z_{\sun}$. \label{tevoldene}}
\end{figure}

\begin{figure}
\epsscale{0.80}
\plotone{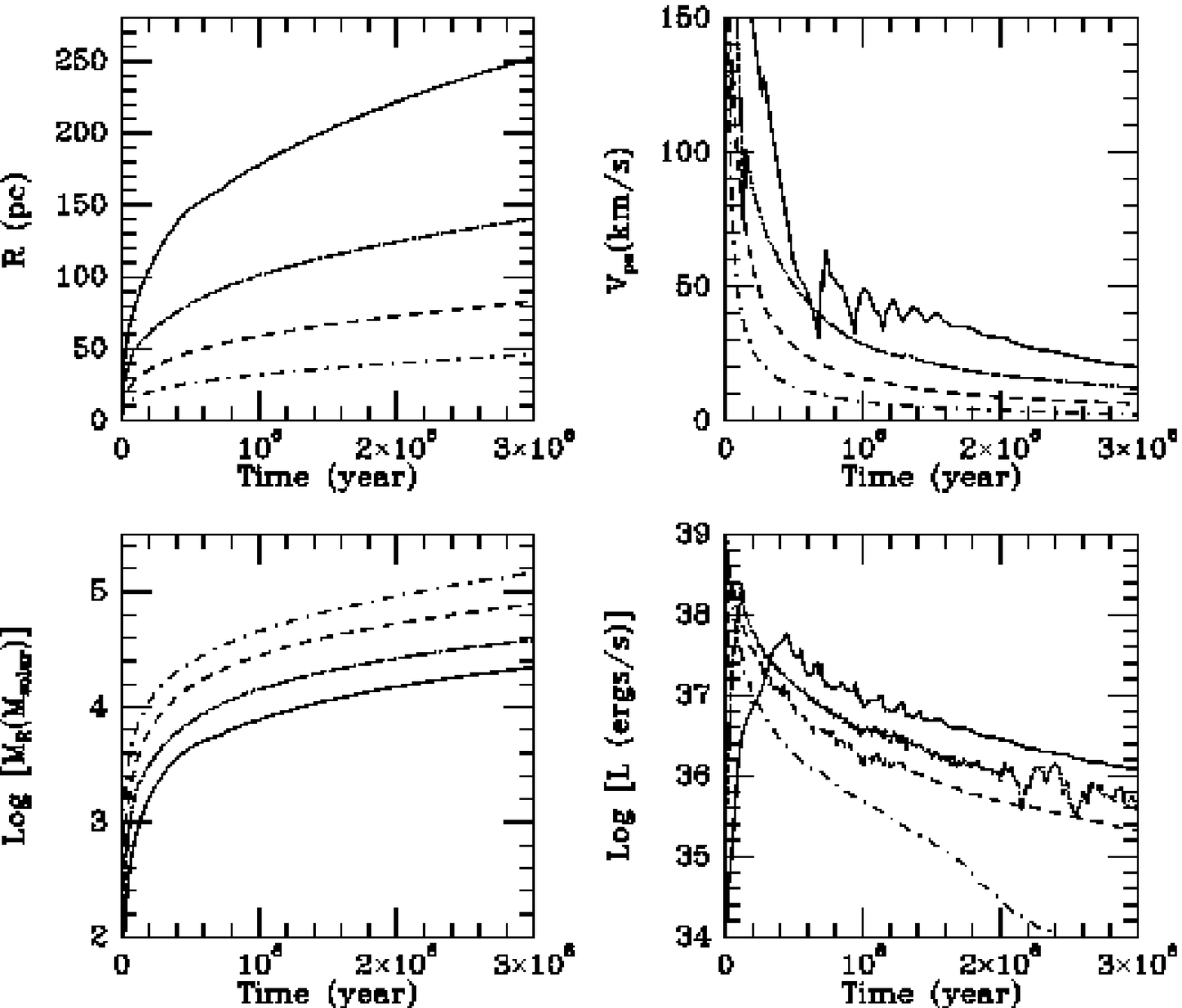}
\caption[f9.eps]{The radius $R$ and the mass
of the SNR $M_{R}$, the post-shock fluid velocity  $V_{ps}$, and the luminosity 
of the SNR $L$,
as a function of time,
for the cases of Fig.~\ref{tevoldene}. \label{tevoldeno}}
\end{figure}

Figures \ref{tevoldene} and \ref{tevoldeno}  
show various global quantities as functions of time for
$Z = 0.02$ and several ISM densities. The total energy, the kinetic 
energy,
and the thermal energy of the SNR ($E_{Rtot}$, $E_{Rkin}$,
and $E_{Rth}$, respectively), 
the total energy and the kinetic energy of the shell ($E_{Stot}$ and
$E_{Skin}$,
respectively), and the thermal energy in the hot
bubble $E_{Bth}$ are plotted against time in Fig.~\ref{tevoldene}.
The radius $R$ and the mass
of the SNR, $M_{R}$, the post-shock fluid velocity, $V_{ps}$, 
and the luminosity
of the SNR, $L$,
as a function of time are plotted in Fig.~\ref{tevoldeno}.
The increase in the total energy at very late
times for the high-density case is due to the accumulation of 
matter and thermal energy from the ambient ISM.  This will be discussed 
in more detail later in this section.  
For all cases, it is clear
that the energy input from a supernova is significantly
less than the initial $10^{51}$ ergs, a value which has been assumed in 
various models of galactic
and globular cluster formation and evolution.  In addition, the 
strong dependence of the SNR evolution on the ambient density should 
be noted.  
This is due to the fact that the cooling rate is proportional to the square of
the local density behind the shock
(which is influenced by the pre-shock density values). 

In our SNR models, 
the time scales of cooling vary from about $10^4$ yr to 
a few $10^5$ 
yr.  These time-scales are smaller than the size 
of time steps
taken in galactic models, due to the fact that the
total evolution time for such models tend to be of the order of 10 Gyr.
The numerical restrictions
typically limit the total number of time steps to 10,000 to 100,000 time
steps, depending on how much computation is involved in each time step.   
As a result, the time steps in such models are limited 
to about $10^5$ yr at most, much larger than the representative
cooling time or dynamical time of SNR evolution.  Therefore, it is 
clear that galactic models cannot
resolve the shocks either in time or in space.

The effects of interactions at the shock front 
with waves created in reflection of the initial reverse shock, 
discussed in \S\ref{ST}, are best visible 
in the lowest density case ($\rho_0 = 0.0133 {\rm m_H}/{\rm cm}^3$, 
solid line in 
Fig.~\ref{tevoldene}).  
The evolution time scale is a steep power of $\rho_0$, and therefore 
more wave interactions and 
details of the early phase can be seen in the lowest density case.
A steep decrease in the post-shock fluid velocity indicates the
deceleration of the shock front due to the pressure gradient, 
as a result of strong cooling.  The density is enhanced further,
resulting in a higher cooling rate.   
Eventually, the wave created by reflection of initial reverse shock
approaches the shock front.
As it reaches the shock front, the interaction increases
the shock velocity, thereby decreasing the density in the front.  This
then results in a sudden decrease of the cooling rate.  The amplitudes
and the velocities
of such waves dissipate with time, as they encounter the shell of 
the contact discontinuity and that of
the shock front.  Some fraction transmits through
the contact discontinuity, travels to the center, and then reflects back, 
again encountering the contact discontinuity.  After a few $t_0$, 
major wave interactions with the shock front are no longer observed. 

The highest density case (dash-dotted line in Fig.~\ref{tevoldene}
and \ref{tevoldeno})
is helpful in illustrating the post-cooling phase.
When the luminosity L becomes two orders of magnitude less than that of 
the peak value, it is clear that the thermal energy of the SNR 
stops decreasing.  It reaches to a minimum, stays
constant, then starts increasing. 
This increase is due to the fact that the cooling rate falls below
the rate of thermal energy accumulation from the surroundings, in
addition to the energy converted from kinetic energy in the shock.
The cooling continues mainly in the hot bubble, but at a much lower
rate.  Cooling in an unshocked medium can be and is currently 
included in models of stellar systems, as they can easily be resolved
both in time and in space.   Therefore, the effects of continuing cooling in
the bubble and the shell are not considered here, but are left for  
models of stellar systems to include as cooling of the ISM.  

The different behavior exhibited by
the various density cases are significant enough
that it is necessary to formulate the SN energy input to the ISM as 
a function of
density.  We will therefore give
a description of SNR properties as a function of the ambient density in  
\S\ref{fits}.

\subsection{The Effect of Metallicity}
\label{metaleffect}
\begin{figure}
\epsscale{0.80}
\plotone{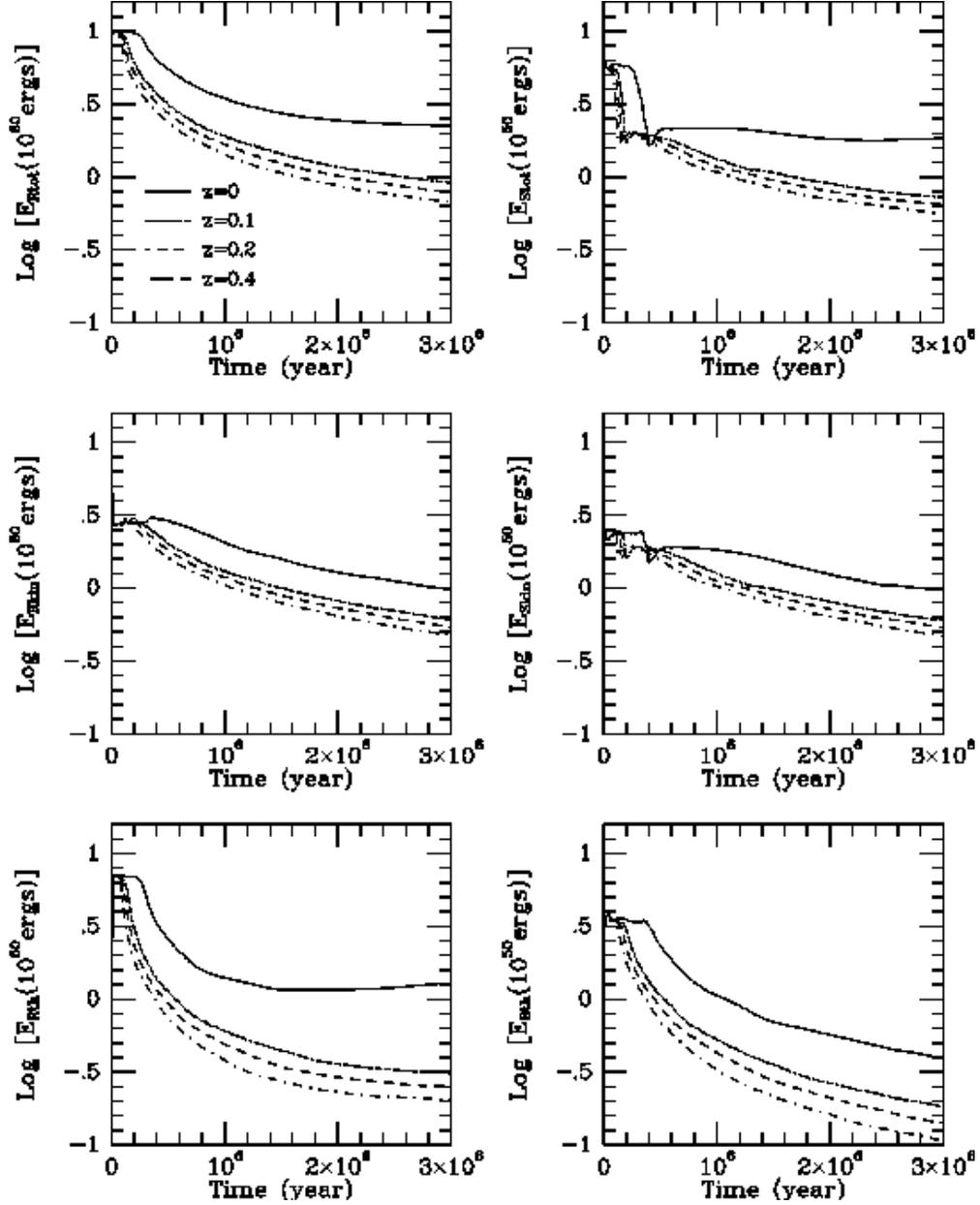}
\caption[f10.eps]{Total energy, kinetic energy,
and thermal energy of the SNR ($E_{Rtot}$, $E_{Rkin}$,
and $E_{Rth}$, respectively) and
total energy, and kinetic energy of the shell ($E_{Stot}$ and
$E_{Skin}$,
respectively), and thermal energy in the hot
bubble $E_{Bth}$ vs. time,
for various cases of metallicity 
($Z = 0.00$ (solid line), $0.01$ (dotted line), $0.02$ (dash line), and
$0.04$ (dash-dot line) in the unit of $m_H/{\rm cm}^3$).  The ambient density 
is fixed at $\rho_0 = 0.133 m_H/{\rm cm}^3$. \label{tevolmete}}
\end{figure}

\begin{figure}
\epsscale{0.80}
\plotone{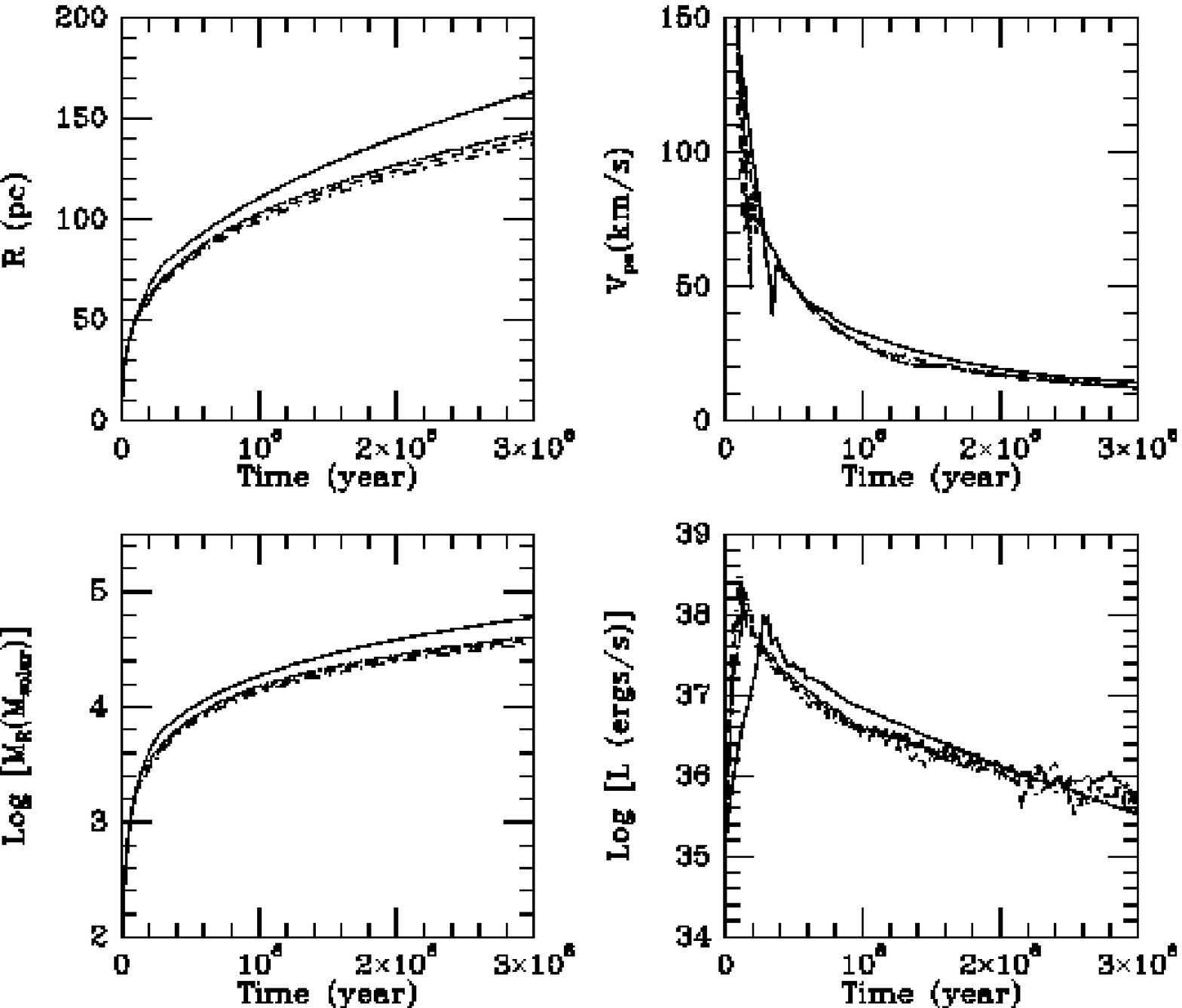}
\caption[f11.eps]{The radius $R$ and the mass
of the SNR $M_{R}$, the post-shock fluid velocity  $V_{ps}$, and the luminosity 
of the SNR $L$,
as a function of time, for the cases in Fig.~\ref{tevolmete}. \label{tevolmeto}}
\end{figure}

Figures \ref{tevolmete} and \ref{tevolmeto} 
show the evolution of various global quantities for 
$\rho_0 = 0.133\, {\rm m_H} /{\rm cm}^3$ and 
$Z = 0.00$, $0.01$, $0.02$, and $0.04$.  The total energy, the 
kinetic energy,
and the thermal energy of the SNR ($E_{Rtot}$, $E_{Rkin}$,
and $E_{Rth}$, respectively), 
total energy and the kinetic energy of the shell ($E_{Stot}$ and
$E_{Skin}$,
respectively), and the thermal energy in the hot
bubble $E_{Bth}$ are plotted against time in Fig.~\ref{tevolmete}.  
The radius $R$ and the mass
of the SNR, $M_{R}$, the post-shock fluid velocity, $V_{ps}$, and the luminosity
of the SNR, $L$,
as a function of time are plotted in Fig.~\ref{tevolmeto}.
The low-metallicity case indeed exhibits
a slower rate of cooling 
and a smaller energy loss.  Nevertheless, the loss is already significant 
after 1 Myr, a short time scale compared to the galactic evolution time scale. 
The neglect of the cooling in the shell
is thus inappropriate,
even for the case of a zero metallicity environment.  The 
differences in energy evolution between the models stem from the 
very efficient metal cooling, which leads to considerably larger values
of the cooling function at temperatures between $10^5$~K and $10^7$~K.
This illustrates the need to explicitly include the dependence of 
the supernova energy input on the environmental metallicity.  The actual
form of this dependence is given in the next section.

The difference is most significant between a case with a moderate
amount of metals and
the case with very low metallicity ($\log{Z/Z_{\sun}} \leq -2$).  
The time scale of evolution clearly depends
on the metallicity.  There is a difference of a factor of about three 
in the time scale between
the low-metallicity case and the solar-metallicity case.  In addition,
more energy is retained if the metallicity is low.  These are the consequences
of metallicity dependent cooling (see Fig.~\ref{coolingfunction}).
The highly efficient cooling by metals results in larger values of 
the cooling function for the higher metallicity cases.  
On the other hand, the cooling in a low metallicity gas is inefficient,
due to the absence of metal cooling.

Note that we have assumed that the cooling functions are calculated with the
metallicity of the ISM, not of the ejecta-ISM mixture.
The validity of this assumption is strongly suggested by the results
of a 2-D calculation (Gaudlitz\markcite{Gaud1996} 1996).
Although a significant amount of mixing occurs, 
the shell of the SNR itself is 
rather stable during
the radiative phase.  The mixing is confined to the bubble, where 
cooling is not efficient due to the low density.  The shell 
consists of freshly accumulated material from the surroundings,
and therefore the cooling function need not to be modified.  
This assumption is examined in detail in \S\ref{assumptions}.

\section{POWER-LAW FITS FOR THE GLOBAL QUANTITIES}
\label{fits}

For a simple incorporation of these results to global modeling of
stellar/ISM system formation and evolution, we need a set of 
descriptions for the global quantities, such as the energies and masses
of the SNR and the shell.  We will now show that the basic dependences
of these quantities on metallicity and density
are well described by power-law fits.  With the use of such power-laws,
it is possible to include the effects of the environment in supernova
energy input in galaxy or globular 
cluster formation models.  For this purpose, we have widened the range
of the densities and the metallicities we explore.  We have selected 
densities ranging
from $1.33 \times 10^{-3} {\rm m_H}/{\rm cm}^3$ to $1.33 \times 10^{3} 
{\rm m_H}/{\rm cm}^3$,
and $\log{Z/Z_{\sun}}$ ranging from $-3$ to $0.5$.  (The values of 
global quantities for $\log{Z/Z_{\sun}} = -3$ is not included in the
fit, and thus not plotted in the figures.)

%\hbox
\begin{center}
\begin{deluxetable}{rrrrrrr}
\tablenum{1}
\tablecaption{Model Results at $t_0$ (Dynamical and General Properties)}
\tablehead{
\colhead{$\log[\displaystyle {Z \over Z_{\sun}}]$} &
\colhead{$\rho/m_H$} & 
\colhead{$t_0$} &
\colhead{$\log(R)$} &
\colhead{$\log(M_{R})$} &
\colhead{$\log(M_{S})$} &
\colhead{$\log(L_{max})$}\\ 
\colhead{} &
\colhead{$(cm^{-3})$} &
\colhead{$(yr)$} &
\colhead{$(pc)$} &
\colhead{$(g)$} &
\colhead{$(g)$} &
\colhead{$(erg/sec)$}
}
\startdata
$-3.0$ & $1.33E-01$ & $2.88E+05$ & $ 75.4$ & $37.090$ & $36.943$ & $38.20$ \nl

$-2.0$ & $1.33E-01$ & $2.82E+05$ & $74.4$ & $37.072$ & $36.930$ & $38.21$ \nl

$-1.5$ & $1.33E-01$ & $2.69E+05$ & $ 73.3$ & $37.052$ & $36.909$ & $38.24$ \nl

$-1.0$ & $1.33E-01$ & $2.34E+05$ & $ 69.1$ & $36.976$ & $36.824$ & $38.16$ \nl

$-0.5$ & $1.33E-01$ & $1.64E+05$ & $ 59.9$ & $36.814$ & $36.675$ & $38.23$ \nl

%$ 0.0$ & $1.33E-01$ & $1.21E+05$ & $ 53.2$ & $36.655$ & $36.527$ & $38.44$ \nl

$ 0.5$ & $1.33E-01$ & $8.87E+04$ & $ 47.9$ & $36.511$ & $36.339$ & $38.71$ \nl

\nl

$ 0.0$ & $1.33E-03$ & $1.43E+06$ & $368.3$ & $37.166$ & $37.003$ & $36.96$ \nl
$ 0.0$ & $1.33E-02$ & $4.22E+05$ & $142.7$ & $36.929$ & $36.751$ & $37.74$ \nl
$ 0.0$ & $1.33E-01$ & $1.22E+05$ & $ 55.8$ & $36.715$ & $36.542$ & $38.48$ \nl
$ 0.0$ & $1.33E+00$ & $3.44E+04$ & $ 21.4$ & $36.473$ & $36.282$ & $39.22$ \nl
$ 0.0$ & $1.33E+01$ & $9.73E+03$ & $  8.2$ & $36.223$ & $36.033$ & $39.79$ \nl
$ 0.0$ & $1.33E+02$ & $3.06E+03$ & $  3.3$ & $36.011$ & $35.838$ & $40.42$ \nl
$ 0.0$ & $1.33E+03$ & $9.57E+02$ & $  1.3$ & $35.798$ & $35.630$ & $40.96$ \nl

\enddata
 \end{deluxetable}
% \end{center}
% 
% \begin{center}
 \begin{deluxetable}{rrrrrrrr}
\tablenum{1---{\it Continued}}
\tablecaption{Model Results at $t_0$ (Energetics)}
\tablehead{
\colhead{$\log[\displaystyle {Z \over Z_{\sun}}]$} &
\colhead{$\rho/m_H$} &
\colhead{$\log(E_{Rtot})$} &
\colhead{$\log(E_{Stot})$} &
\colhead{$\log(E_{Rkin})$} &
\colhead{$\log(E_{Skin})$} &
\colhead{$\log(E_{Rth})$} &
\colhead{$\log(E_{Bth})$} \\
\colhead{} & 
\colhead{$(cm^{-3})$} &
\colhead{$(erg)$} &
\colhead{$(erg)$} &
\colhead{$(erg)$} &
\colhead{$(erg)$} &
\colhead{$(erg)$} &
\colhead{$(erg)$}
}
\startdata
$-3.0$ & $1.33E-01$ & $50.907$ & $50.605$ & $50.463$ & $50.384$ & $50.713$ & $50.552$ \nl
$-2.0$ & $1.33E-01$ & $50.894$ & $50.588$ & $50.456$ & $50.376$ & $50.697$ & $50.541$ \nl
$-1.5$ & $1.33E-01$ & $50.888$ & $50.573$ & $50.462$ & $50.378$ & $50.683$ & $50.541$ \nl
$-1.0$ & $1.33E-01$ & $50.866$ & $50.529$ & $50.454$ & $50.357$ & $50.654$ & $50.532$ \nl
$-0.5$ & $1.33E-01$ & $50.896$ & $50.603$ & $50.434$ & $50.345$ & $50.713$ & $50.528$ \nl
%$0.0$ & $1.33E-01$ & $50.864$ & $50.573$ & $50.412$ & $50.320$ & $50.675$ & $50.489$ \nl
$0.5$ & $1.33E-01$ & $50.857$ & $50.491$ & $50.416$ & $50.293$ & $50.662$ & $50.539$ \nl

\nl

$ 0.0$ & $1.33E-03$ & $50.923$ & $50.610$ & $50.421$ & $50.295$ & $50.759$ & $50.561$ \nl
$ 0.0$ & $1.33E-02$ & $50.891$ & $50.535$ & $50.422$ & $50.294$ & $50.711$ & $50.565$ \nl
$ 0.0$ & $1.33E-01$ & $50.870$ & $50.562$ & $50.418$ & $50.332$ & $50.681$ & $50.518$ \nl
$ 0.0$ & $1.33E+00$ & $50.852$ & $50.523$ & $50.426$ & $50.336$ & $50.647$ & $50.515$ \nl
$ 0.0$ & $1.33E+01$ & $50.884$ & $50.588$ & $50.438$ & $50.368$ & $50.691$ & $50.528$ \nl
$ 0.0$ & $1.33E+02$ & $50.904$ & $50.633$ & $50.436$ & $50.371$ & $50.723$ & $50.523$ \nl
$ 0.0$ & $1.33E+03$ & $50.888$ & $50.641$ & $50.428$ & $50.363$ & $50.703$ & $50.474$ \nl

\enddata
\end{deluxetable}
%\end{center}

%\begin{center}
\begin{deluxetable}{rrrrrrrr}
\tablenum{2}
\tablecaption{Model Results at $t_f$ (Dynamical and General Properties)}
\tablehead{
\colhead{$\log[\displaystyle {Z \over Z_{\sun}}]$} &
\colhead{$\rho/m_H$} & 
\colhead{$t_f$} &
\colhead{$\log(R)$} &
\colhead{$\log(M_{R})$} &
\colhead{$\log(M_{S})$} &
\colhead{$\log(L_{f})$} &
\colhead{$\log(\displaystyle{L_f \over L_{max}})$}\\
\colhead{} &
\colhead{$(cm^{-3})$} &
\colhead{$(yr)$} &
\colhead{$(pc)$} &
\colhead{$(g)$} &
\colhead{$(g)$} &
\colhead{$(erg/sec)$} &
\colhead{}
}
\startdata

$-3.0$ & $1.33E-01$ & $3.74E+06$ & $175.6$ & $38.181$ & $38.166$ & $35.35$ & $-2.85$ \nl
$-2.0$ & $1.33E-01$ & $3.67E+06$ & $171.2$ & $38.152$ & $38.137$ & $35.85$ & $-2.36$ \nl
$-1.5$ & $1.33E-01$ & $3.50E+06$ & $163.5$ & $38.089$ & $38.075$ & $35.91$ & $-2.33$ \nl
$-1.0$ & $1.33E-01$ & $3.04E+06$ & $150.3$ & $37.992$ & $37.981$ & $35.95$ & $-2.21$ \nl
$-0.5$ & $1.33E-01$ & $2.13E+06$ & $129.5$ & $37.812$ & $37.800$ & $36.14$ & $-2.09$ \nl
%$ 0.0$ & $1.33E-01$ & $1.57E+06$ & $112.9$ & $37.671$ & $37.659$ & $36.21$ & $-2.23$ \nl
$ 0.5$ & $1.33E-01$ & $1.15E+06$ & $ 99.5$ & $37.483$ & $37.470$ & $36.31$ & $-2.40$ \nl

\nl

$ 0.0$ & $1.33E-03$ & $1.86E+07$ & $803.4$ & $38.183$ & $38.158$ & $35.16$ & $-1.80$ \nl
$ 0.0$ & $1.33E-02$ & $5.49E+06$ & $301.6$ & $37.916$ & $37.900$ & $35.70$ & $-2.04$ \nl
$ 0.0$ & $1.33E-01$ & $1.59E+06$ & $114.3$ & $37.647$ & $37.630$ & $36.18$ & $-2.30$ \nl
$ 0.0$ & $1.33E+00$ & $4.47E+05$ & $ 43.0$ & $37.386$ & $37.371$ & $36.63$ & $-2.59$ \nl
$ 0.0$ & $1.33E+01$ & $1.26E+05$ & $ 16.4$ & $37.147$ & $37.131$ & $37.19$ & $-2.60$ \nl
$ 0.0$ & $1.33E+02$ & $3.98E+04$ & $  6.6$ & $36.923$ & $36.907$ & $37.91$ & $-2.51$ \nl
$ 0.0$ & $1.33E+03$ & $1.24E+04$ & $  2.5$ & $36.793$ & $36.783$ & $38.26$ & $-2.70$ \nl

\enddata
\end{deluxetable}
\end{center}

\begin{center}
\begin{deluxetable}{rrrrrrrr}
\tablenum{2---{\it Continued}}
\tablecaption{Model Results at $t_f$ (Energetics)}
\tablehead{
\colhead{$\log[\displaystyle {Z \over Z_{\sun}}]$} &
\colhead{$\rho/m_H$} &
\colhead{$\log(E_{Rtot})$} &
\colhead{$\log(E_{Stot})$} &
\colhead{$\log(E_{Rkin})$} &
\colhead{$\log(E_{Skin})$} &
\colhead{$\log(E_{Rth})$} &
\colhead{$\log(E_{Bth})$} \\
\colhead{} & 
\colhead{$(cm^{-3})$} &
\colhead{$(erg)$} &
\colhead{$(erg)$} &
\colhead{$(erg)$} &
\colhead{$(erg)$} &
\colhead{$(erg)$} &
\colhead{$(erg)$}
}
\startdata
$-3.0$ & $1.33E-01$ & $50.332$ & $50.262$ & $49.921$ & $49.917$ & $50.118$ & $49.488$ \nl
$-2.0$ & $1.33E-01$ & $50.266$ & $50.187$ & $49.924$ & $49.919$ & $50.002$ & $49.472$ \nl
$-1.5$ & $1.33E-01$ & $50.159$ & $50.059$ & $49.904$ & $49.896$ & $49.808$ & $49.457$ \nl
$-1.0$ & $1.33E-01$ & $50.119$ & $50.017$ & $49.924$ & $49.920$ & $49.678$ & $49.427$ \nl
$-0.5$ & $1.33E-01$ & $50.117$ & $50.006$ & $49.952$ & $49.947$ & $49.616$ & $49.454$ \nl
%$ 0.0$ & $1.33E-01$ & $50.111$ & $50.023$ & $49.986$ & $49.984$ & $49.510$ & $49.370$ \nl
$ 0.5$ & $1.33E-01$ & $50.035$ & $49.950$ & $49.924$ & $49.921$ & $49.389$ & $49.273$ \nl

\nl

$ 0.0$ & $1.33E-03$ & $50.242$ & $50.105$ & $49.987$ & $49.968$ & $49.890$ & $49.635$ \nl
$ 0.0$ & $1.33E-02$ & $50.135$ & $50.024$ & $49.958$ & $49.954$ & $49.658$ & $49.472$ \nl
$ 0.0$ & $1.33E-01$ & $50.040$ & $49.932$ & $49.903$ & $49.888$ & $49.473$ & $49.330$ \nl
$ 0.0$ & $1.33E+00$ & $49.967$ & $49.905$ & $49.881$ & $49.877$ & $49.222$ & $49.067$ \nl
$ 0.0$ & $1.33E+01$ & $49.972$ & $49.939$ & $49.924$ & $49.923$ & $48.988$ & $48.807$ \nl
$ 0.0$ & $1.33E+02$ & $49.915$ & $49.894$ & $49.883$ & $49.882$ & $48.755$ & $48.550$ \nl
$ 0.0$ & $1.33E+03$ & $50.048$ & $50.041$ & $50.035$ & $50.034$ & $48.506$ & $48.177$ \nl

\enddata
\end{deluxetable}
\end{center}

In order to obtain a realistic measure of 
the cooling in the shell created by shocks, the final
model was taken at $t_f = 13 \times t_{0}$, where $t_{0}$ is the age of 
SNR at the maximum luminosity (or maximum cooling rate).  
This value was chosen such that the 
models have a luminosity of approximately 0.5\% of the maximum
value by this time.  The cooling continues slowly; such effect must be
taken into account separately as a cooling of the (unshocked) ISM in any
application of our results. 

\begin{figure}
\epsscale{0.80}
\plotone{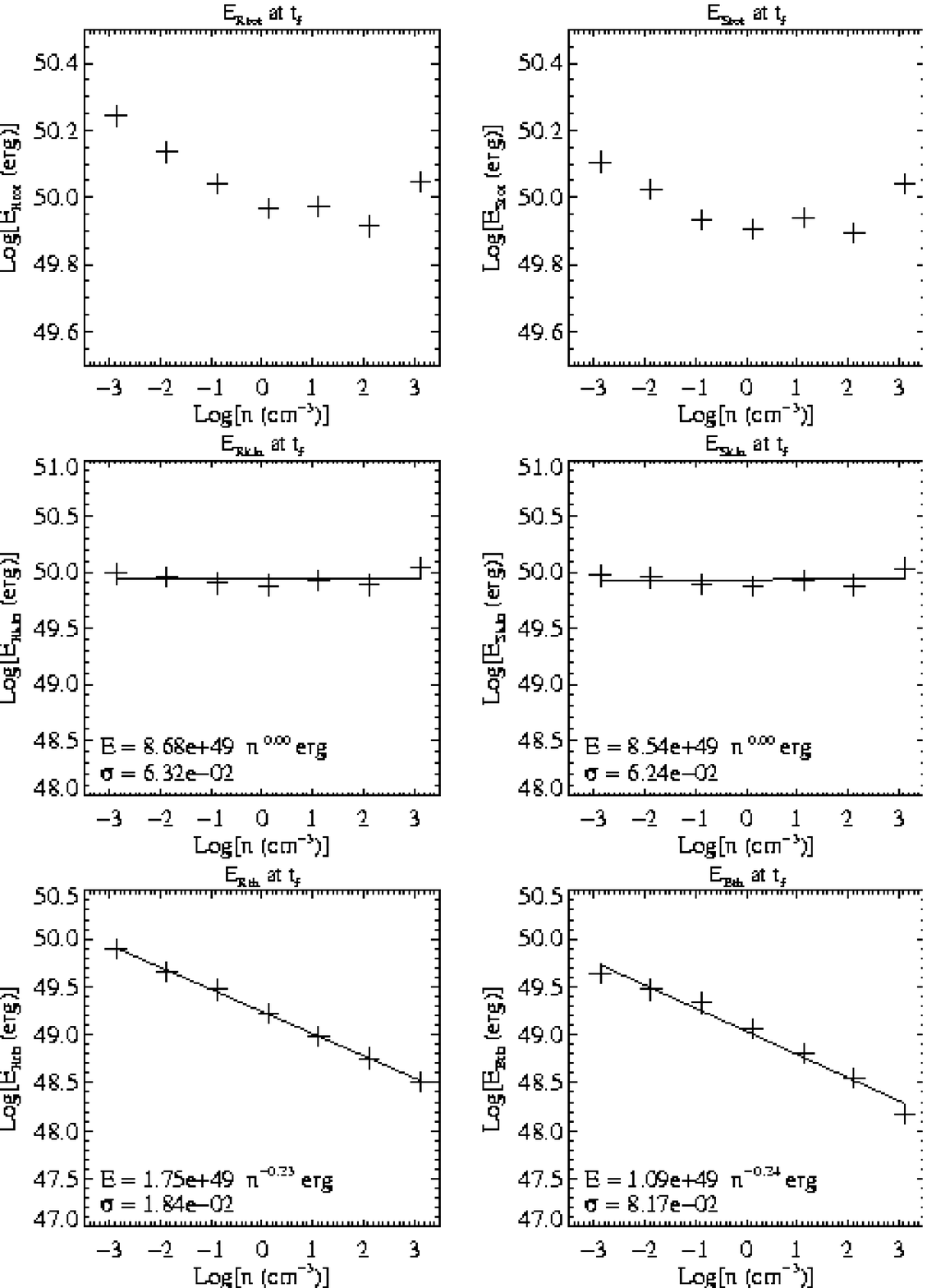}
\caption[f12.eps]
{The global quantities relevant for prescribing the effects of 
SN explosions on the surrounding ISM as a function of
the ambient density, and the least-square fits 
are shown where appropriate.  The quantities are total energy, kinetic energy,
and thermal energy of the SNR ($E_{Rtot}$, $E_{Rkin}$,
and $E_{Rth}$, respectively) and
total energy, and kinetic energy of the shell ($E_{Stot}$ and
$E_{Skin}$,
respectively), and thermal energy in the hot
bubble $E_{Bth}$. 
All quantities are taken at $t_{f}$,
defined by $13 \times t_0$, where $t_{0}$ is the time the maximum luminosity 
is attained. 
The metallicity
is fixed at $Z_{\sun}$.  $\sigma$ is the standard deviation.
\label{fitdene}}
\end{figure}

\begin{figure}
\epsscale{0.80}
\plotone{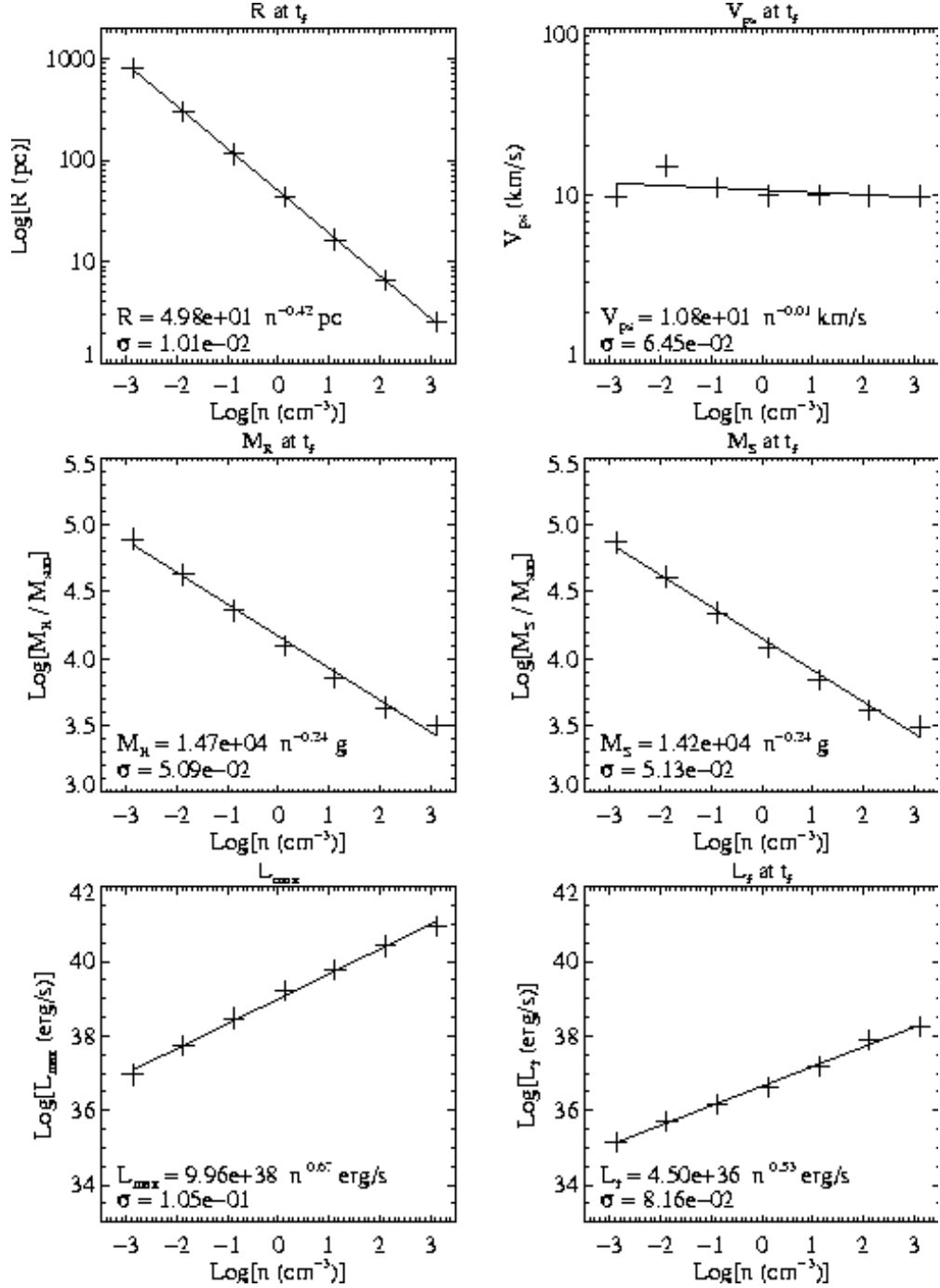}
\caption[f13.eps]
{Similar to Fig.~\ref{fitdene}.  Quantities plotted and fitted are
the radius $R$, the total mass $M_{R}$,
the maximum
luminosity attained $L_{max}$, the post-shock fluid velocity $V_{ps}$,
the shell mass $M_{S}$, and
the luminosity $L_{f}$. 
All quantities except for $L_{max}$ are taken at $t_{f}$ (See 
Fig.~\ref{fitdene}).
The metallicity
is fixed at $Z_{\sun}$.
\label{fitdeno}}
\end{figure}

\begin{figure}
\epsscale{0.80}
\plotone{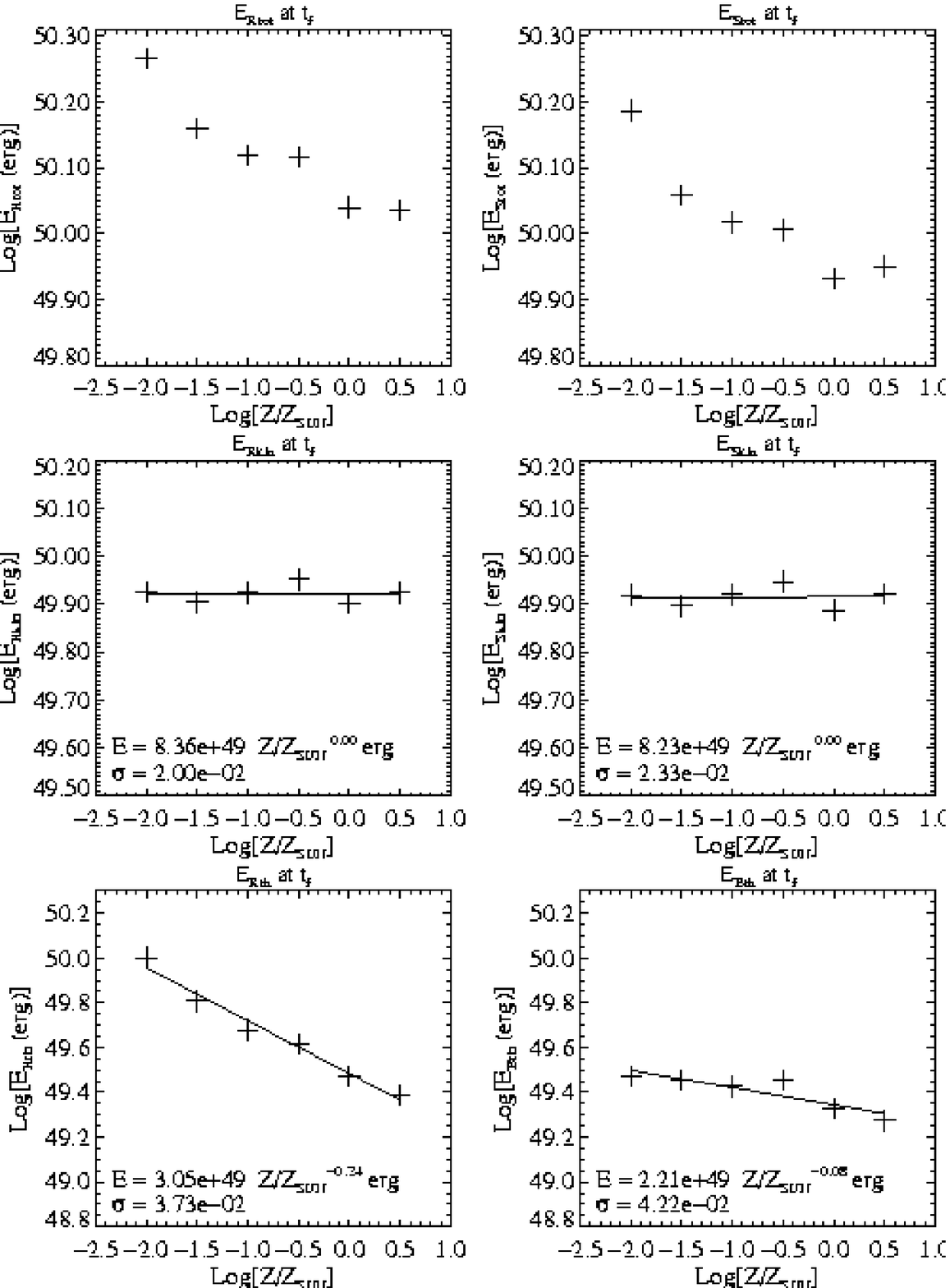}
\caption[f14.eps]{Same as in Fig.~\ref{fitdene}, but 
they are plotted as a function of the ambient metallicity $Z/Z_{\sun}$. 
The ambient density
is fixed at $\rho_0 = 0.133 m_H/{\rm cm}^3$.
\label{fitmete}}
\end{figure}

\begin{figure}
\epsscale{0.80}
\plotone{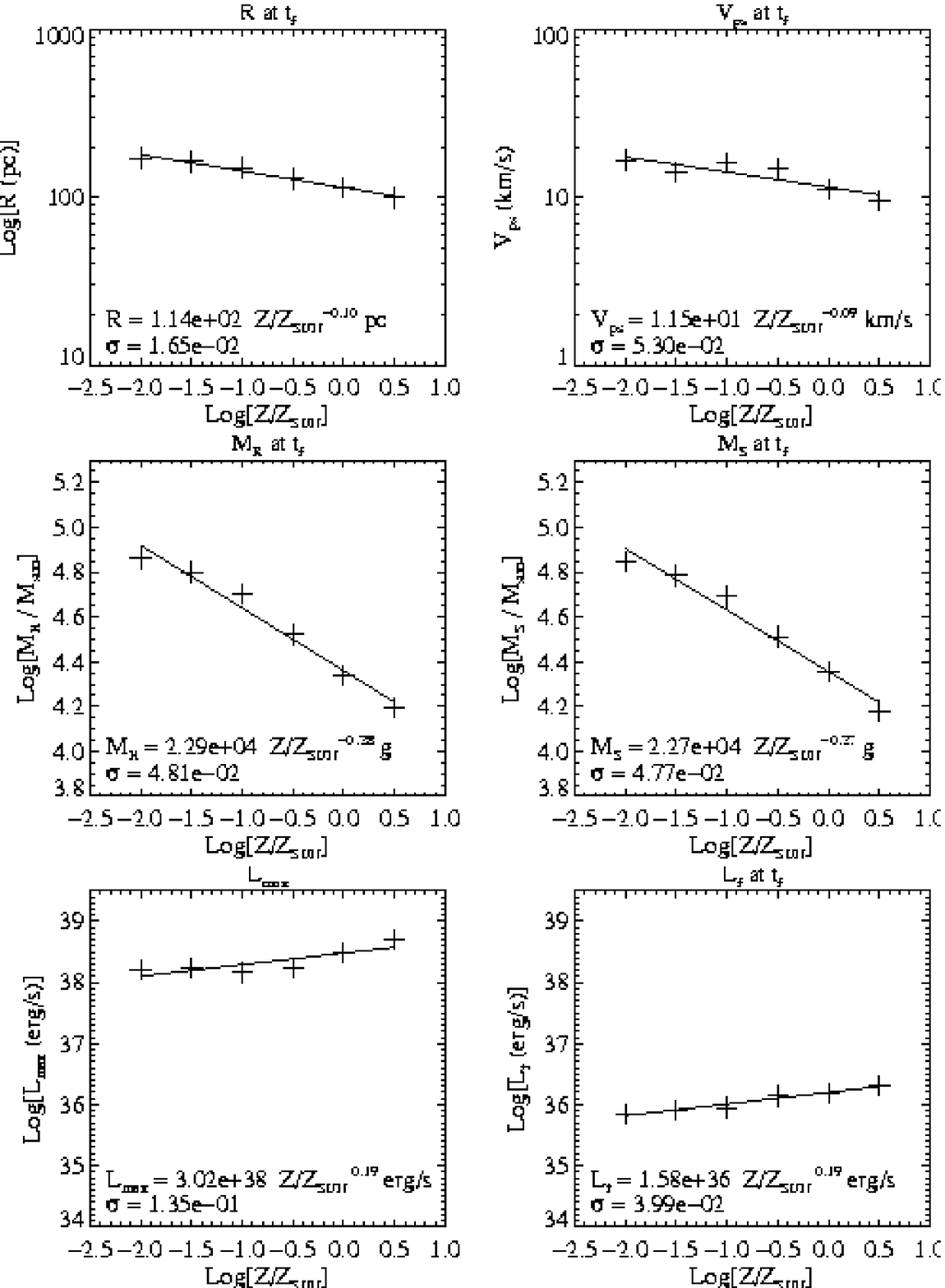}
\caption[f15.eps]{Same as in Fig.~\ref{fitdeno}, but they are 
plotted as a function of the ambient metallicity $Z/Z_{\sun}$. 
The ambient density
is fixed at $\rho_0 = 0.133 m_H/{\rm cm}^3$.
\label{fitmeto}}
\end{figure}

Based upon the results from our numerical calculations for SNRs in different
density and metallicity environments, we can now identify the dependences
of critical quantities on those parameters.
The results are presented in two ways.  First, Table 1 and 2 contain all of the
values of global properties at $t_{0}$ and $t_{f}$, respectively.  
Second, power-law
fits were constructed for the results.
Figures \ref{fitdene} through \ref{fitmeto} illustrate the fits obtained.

The energies of the early supernova remnant evolution
are almost constant across the range
of environmental density and metallicity we have explored,
as seen in Table 1.  This indicates
that the SN evolves almost adiabatically until about $t_0$.
The metallicity and density dependent cooling
has not affected the SNRs to this stage.
On the other hand, the dependences of cooling on metallicity and density
are clearly seen in 
Table 2, taken at $t_f$.  For this table, the number of significant
digits is determined so that the difference between SNR quantities (e.g.,
$M_R$) and shell (or bubble) quantities (e.g., $M_S$) is a well-
determined number.  Those differences can be very small, and
keeping fewer digits would have yielded zero, due to rounding.
The number of the digits does not reflect the accuracy, but they are chosen 
for practical purposes.  

The fit with metallicity was split into two parts because of
the nonlinear dependence of cooling on metallicity.
Near $\log (Z/Z_{\sun}) = -2$ and below, the cooling is dominated by
hydrogen and helium, and therefore the strong dependence of the cooling
efficiency on the metallicity disappears (see Fig.~\ref{coolingfunction}).
For $\log (Z/Z_{\sun}) >
-2$, the strong metallicity dependence of the cooling efficiency
manifests itself in almost all global quantities plotted in Figures 
\ref{fitmete} and \ref{fitmeto} .

The fits for all quantities of interest at $t_{f}$ are presented for 
$\log[Z/Z_{\sun}] > -2$ in the first column below.  
For $\log[Z/Z_{\sun}] \leq -2$, the metallicity dependence disappears;
the corresponding fits are given in the second column.

\begin{minipage}[h]{2.9in}
$${\log[Z/Z_{\sun}] > -2}$$
$$E_{Rtot} = E_{Rkin} + E_{Rth}$$
$$E_{Rkin} = 8.52 \times 10^{49} E_{51} {\rm ergs}$$
$$E_{Rth} = 1.83 \times 10^{49} E_{51} n_0^{-0.23} (Z/Z_{\sun})^{-0.24} 
{\rm ergs}$$
$$E_{Stot} = E_{Skin} + (E_{Rth} - E_{Bth})$$
$$E_{Skin} = 8.39 \times 10^{49} E_{51} {\rm ergs}$$
$$E_{Bth} = 1.23 \times 10^{49} E_{51} n_0^{-0.24} (Z/Z_{\sun})^{-0.08}
{\rm ergs}$$
$$R = 49.3 E_{51}^{2/7} n_0^{-0.42} (Z/Z_{\sun})^{-0.1} pc$$
$$M_{R} = 1.44 \times 10^4 E_{51}^{6/7} n_0^{-0.24} (Z/Z_{\sun})^{-0.28}
   M_{\sun}$$
$$M_{S} = 1.41 \times 10^4 E_{51}^{6/7} n_0^{-0.24} (Z/Z_{\sun})^{-0.27}
   M_{\sun}$$
$$V_{s} = 11.3 E_{51}^{1/14} n_0^{-0.01} (Z/Z_{\sun})^{-0.09} km/s$$
$$L = 4.55 \times 10^{36} E_{51}^{11/14} n_0^{0.53} (Z/Z_{\sun})^{0.19}
erg/s$$
\end{minipage}
%For $\log[Z/Z_{\sun}] \leq -2$, the metallicity dependence disappear, and
%the quantities are given by: 
%\hspace{3in}
\begin{minipage}[h]{2.9in}
$${\log[Z/Z_{\sun}]} \leq -2$$
$$E_{Rtot} = E_{Rkin} + E_{Rth}$$
$$E_{Rkin} = 8.52 \times 10^{49} E_{51} {\rm ergs}$$
$$E_{Rth} = 5.53 \times 10^{49} E_{51} n_0^{-0.23} {\rm ergs}$$
$$E_{Stot} = E_{Skin} + (E_{Rth} - E_{Bth})$$
$$E_{Skin} = 8.39 \times 10^{49} E_{51} {\rm ergs}$$
$$E_{Bth} = 1.78 \times 10^{49} E_{51} n_0^{-0.24} {\rm ergs}$$
$$R = 78.1 E_{51}^{2/7} n_0^{-0.42} pc$$
$$M_{R} = 5.23 \times 10^4 E_{51}^{6/7} n_0^{-0.24} M_{\sun}$$
$$M_{S} = 4.89 \times 10^4 E_{51}^{6/7} n_0^{-0.24} M_{\sun}$$
$$V_{s} = 17.1 E_{51}^{1/14} n_0^{-0.01} km/s$$
$$L = 1.90 \times 10^{36} E_{51}^{11/14} n_0^{0.53} {\rm ergs}/{\rm s}$$
\end{minipage}

In these expressions,
$E_{51}$ is the initial explosion energy in $10^{51}$ ergs, and 
$n_0$ is defined by $\rho_0/{m_H}$.
Other quantities are as defined earlier.  
The initial energy dependence was determined from test runs, and  
was found to be consistent with 
the existing studies of CMB and others.
The validity of such solutions comes from the fact that the dynamical
state of the final stage is still dominated by the pressure-driven
snowplow phase of 
evolution, as the final time marks approximately the end of the 
pressure-driven snowplow phase.  
For this reason, we have adopted the exponents from the
previous studies, which are very close to our numerical results.  A 
simple analysis shows that the 
exponent of $E_{51}$ is a slowly varying function of the metallicity
(Cioffi \& Shull 1991),
but we will ignore this effect here since it is small compared to
other uncertainties involved. 

There is an upward systematic error in the energy, which increases
as the density decreases.  The source is the thermal energy contributed by
the ambient medium.  For the worst case (i.e., the lowest density case), 
this error is estimated to
be about 3\% of the final total energy.  

It should be cautioned that the results so far are purely empirical.
The dependences of these quantities on $n_0$ and $Z$ have no analytical
bases.  Therefore, any extrapolation of the results into regions 
of parameter space beyond that which we have explored in this paper
should be done with caution. 
\section{DISCUSSION}
\label{discussion}
\subsection{Validity and Possible Consequences of Assumptions}
\label{assumptions}

We have made every effort throughout this study of supernova remnant
evolution to insure that we have employed the best available input physics
and that our numerical results are accurate.  In this section, we comment 
briefly on the assumptions we have made and the constraints they might
impose on the applicability of our results.

We did not include the effects of magnetic fields,
as mentioned in \S\ref{methods}.
This assumption
is justified if the Alfv\'en speed is negligible compared to the
shock velocity, or in other words, if the post-shock pressure
is much greater than the magnetic pressure.  Whether this
condition is met or not depends upon the strength of
the magnetic fields in the ISM.  Currently, we do not know how
magnetic fields are created in the Universe, and how
the strength evolves during the lifetime of galaxies. Therefore, we
have chosen not to include magnetic fields in these calculations.
Examples of effects of
magnetic fields are
slower expansion, a smaller final radius, and
less energy loss (due to less compression of the shell); that
is, the SNR influenced by the (random) magnetic fields of
its surroundings would keep more energy, but stay compact
(Slavin \& Cox\markcite{Slav1992} 1992).
It should be noted that the magnetic fields would play
a role to some extent in the evolution of SNRs in the present-day solar
environment, with its estimated field strength of about $5 \mu G$,
as suggested by the authors.   

We have also ignored the effects of turbulence 
by assuming spherical 
symmetry of the SNR.  
However, signatures of instabilities are seen in observed 
young SNRs (Bartel, {\it et~al.}\markcite{Bart1991} 1987, 1991; Wilkinson
\& de Bruyn\markcite{Wilk1990} 1990).  
The stability
of the thin shell has been questioned and studied by others 
(Gull\markcite{Gull1973} 1973; Chevalier, Blondin, \& 
Emmering\markcite{Chev1992} 1992; 
Chevalier \& Blondin\markcite{Chev1995} 1995).  
Here, we merely discuss the consequences of 
non-sphericity with respect to the global dynamics and energetics
of SNRs. 

Turbulence is important in two aspects.  First, the shell, which 
is assumed to be stable, may become unstable and change the dynamical
properties of the SNR.  This may, as a result, change the cooling history
of the SNR, and therefore the energetics.  Secondly, turbulence 
facilitates mixing, which brings the metal rich ejecta into the 
material which was accreted from the surroundings.  
The cooling rate would change only if the turbulence
were to carry a significant amount of metals into the shell. 

As it was pointed out in \S\ref{metaleffect}, 
a 2-D hydrodynamic code was used by Gaudlitz\markcite{Gaud1996}
(1996) to calculate 
the evolution of SNRs similar to the ones we considered here. 
Their calculation showed that mixing was confined
to the bubble, and that the shell is stable for the time scales of 
interest in our study.
The validity of this result relies on whether the spherically 
symmetric initial condition assumed in the model is satisfactory.  In reality, 
turbulence would have already been established in the SNR and in its 
shell by the age the model is initially started.  It is difficult
to predict how turbulence in the very young SNR influence the
subsequent evolution.  In addition, the stability of the shell changes 
as its structure changes dynamically.
This may be an issue which needs more attention
in the future.

We believe that significant mixing of metals into the shell is
quite unlikely except at the very early and late phases, independent of the 2-D 
results.  
The radial velocity profiles in various stages of SNR evolution 
suggest that the ejecta 
enriched material has substantially less radial velocity than the
shell, making
it difficult for the enriched material to catch up with it.
In the ejecta-dominated phase, we expect mixing to occur easily 
because there is not a large layer of accreted ambient matter 
between the ejecta and the shell.  We also expect a Rayleigh-Taylor 
instability due to the density and velocity structure.
In the 
very late phases, the shock velocity can become quite small, and
more efficient mixing may take place.  However, the density
in the extended bubble is much less than that of the thin shell,
and therefore, the variation in the cooling rate would not change
the dynamics nor the energetics noticeably.  In any case, our final
shell characteristics
are extracted before the models reach this phase, and therefore
it would not make a significant 
difference in the results presented in \S\ref{fits}.

We have ignored the effects of thermal conduction.
In the figures of the structures (Fig.~\ref{str9810} though \ref{str1.52e6}),
the temperature profile shows non-uniform structure inside the remnant.
In reality, thermal conduction will smooth the profile, keeping it
approximately uniform in the inner region, away from the shell
(Chevalier\markcite{Chev1975} 1975; Solinger, Buff, \&
Rappaport\markcite{Soli1975} 1975).  This
does not change the overall behavior of the remnant significantly,
since the dominant dynamical and cooling processes occur in and near
the shell of the remnant. 

In addition, the contact discontinuity seen at r = 8 pc in
Fig.~\ref{str9810} ($t = 9810$ yr) and
at r = 15 pc in Fig.~\ref{str1.27e5}
($t = 1.27 \times 10^5$ yr)
is smeared by the effects of thermal conduction, as is the temperature
profile.
Although the
higher density at the contact discontinuity causes extra cooling, it is
more than two orders of magnitude less than the value at the shock front,
and therefore, does not affect the global cooling history.  Therefore,
the global characteristics of the SNR are not significantly affected
by the approximation to ignore thermal conduction.

\subsection{Implications of the Results}
\label{implications}
The major implication of our results is that the assumptions made in
incorporating the energy input from SN explosions to galaxy/globular
cluster formation and evolution models should be reconsidered.  First,
the value of energy input per SN is often overestimated by 
a factor of about 10.  Secondly, the ratio between the amounts
that become kinetic energy and thermal energy is not correctly 
estimated, since the values are often determined from a phase of 
the SNR that is too early in its evolution.  Finally, the effect
of the ambient medium on the SNR evolution, which influences the
above quantities, is not taken into account.  In addition, other
effects of SNe on the ISM, such as the production of clouds 
in the shell, are not taken into account consistently.  

We will now give a few specific examples from existing studies of galactic
formation/evolution.

Chemodynamical models (Theis, Burkert \& Hensler\markcite{Thei1992} 1992, 
Burkert, Hensler \& Truran\markcite{Burk1992} 1992) combine dynamical 
modeling of 
galaxies with microphysics of the ISM and effects of star formation and 
evolution to produce results which predict the dynamical state 
of a galaxy, as well as the chemical compositions.  It is a powerful
method in studying the history of the Galaxy, giving considerably 
more information than the studies of dynamics or chemistry separately.
The results typically include the chemical compositions and kinematical
information as functions 
of location.  Given observational data for comparisons
to restrict the model parameters or assumptions, they provide 
a significantly more reliable history of formation and
evolution of galaxies.

Because of its complex nature, there are several simplifying assumptions
one must make in such modeling.  To treat SN explosions without 
resolving the shocks, one must assume how much energy is provided, where 
(e.g., in the cloud or in the hot medium) and in what form
(kinetic or thermal).
In those studies, 
it was assumed that the SN explosion provides $10^{51}$ ergs. 
Unfortunately, the assumed values of the energy input from each SN were
overestimated in some studies because they were taken 
to be equal to the typical total energy released from 
a SN.  In Theis {\it et~al.}\markcite{Thei1992} 
(1992), it was indicated
that a factor of five change in the value of $E_{SN}$, the amount of energy
input from a SN to the ISM, 
significantly changes the kinematics.
As an example, the resulting velocity dispersion of low mass stars 
varied from 9 km/s ($E_{SN} = 10^{51} 
{\rm ergs}$) to 78 km/s ($E_{SN} = 5 \times
10^{51} {\rm ergs}$).  
The resolution was on a much coarser scale
to accommodate the galactic scale, and therefore SN shocks were not resolved.
As a result, the cooling in the shell of a SNR was not taken
into account, and too much SN
energy was put into the ISM. 

Cole {\it et~al.}\markcite{Cole1994} (1994) performed a extensive 
search in the parameter space of
galactic models
by observational fits.  One of the parameters they examined was
the fraction of supernova energy input that is in the form of 
kinetic energy, $f_v$ (assuming each supernova gives out $10^{51} \rm{ergs}$
in total energy). 
They concluded that the supernova feedback
has a significant influence in their results if $f_v$ is of the order of 0.1.  
They give a best fit value of 0.2, although the model results for $f_v = 0.1$
and for $f_v = 0.2$ differ only slightly.  Our results indicate a value
of about 0.09, which agrees with their conclusion.

In other studies, the value for the supernova energy input is assumed to be
a certain number, or related, usually by a constant factor, to the
energy input from stellar winds (e.g., Rosen \& Bregman\markcite{
Rose1996} 1996).  In reality, 
those values depend 
largely on the star formation history.  In addition, 
the resolution in their model
was better than those in 
the models mentioned earlier, but it is still not sufficient
to resolve the thin shell created in the radiative shock.  
Thus, such a treatment needs
further refinement.  

For studies which concern the formation of galaxies or their evolution
over a large range of age and metallicity, the treatment of
supernovae should include 
the dependence on the metallicity.  No study so far took this effect into
account, as well as the dependence on the ambient density.
In order to 
make models of galaxies robust, the dependences need
to be considered.

\section{CONCLUSIONS}
\label{conclusions}

%***************************************************************************
The significant conclusions to be drawn from the numerical studies presented 
in this paper are the following:
\begin{enumerate}
%***************************************************************************
\item The value of supernova energy input in 
standard assumptions made for the 
incorporation of the SN energy input into models for the 
evolution of galaxies is often overestimated. 
It  commonly assumed that 
the energy input is comparable to the $\approx$ $10^{51}$ ergs
associated with the light curve and kinetic energy output of 
both Type Ia and Type II supernovae. We took this number as the
total initial (kinetic plus thermal)  
energy in our SNR models, and found that much of the energy is lost
in radiation. 
The results are summarized in  
Table 2 for the energetics of the remnants in the late stages of their 
expansion.  The total energies range from $\approx 9 \times 10^{49}$ to
$\approx 3 \times 10^{50}$ ergs, with a typical case being $\approx$ 10$^{50}$
erg, approximately 10\% of the initial total energy.

%***************************************************************************
\item The amount of supernova energy input is a sensitive function of 
the characteristics -- the density and the metallicity -- of the environment. 
The basic dependences are again evident from the model results presented 
in Table 2. The general trends in these cases are relatively straightforward 
to understand. The total energy available in kinetic and thermal energy 
is greatest in the limits of low density and low metallicity. This is a 
direct consequence of the lower cooling rates that occur in these limits 
such that a greater fraction of the initial supernova input energy remains 
in the remnant. 

%***************************************************************************
\item A proper treatment of the problem permitting a realistic 
measure of the relative amounts of thermal energy and kinetic energy is 
important.  The bulk of the supernova energy input provides the kinetic 
energy of cloud motion.  The kinematical properties of clouds are 
directly related to the kinematical characteristics of stars formed in these
clouds.   
On the other hand, the fraction of the energy  
in the bubble keeps the gas hot for the time scale of interest.
Therefore, mishandling the relative energy input may cause 
an overestimation of the thermal energy input and an underestimation 
of the kinetic energy input, or vise versa,
changing the model predictions.

%***************************************************************************
\item A proper treatment of the problem permitting a realistic estimate
of the relative amounts of shell (cloud) energy and bubble 
(hot gas) energy is important.  As mentioned above, the proper energy
divisions into kinetic and thermal energy is critical.  Since many galactic
models distinguish a cloudy medium and hot medium, the 
proper method for distributing the total energy input from supernovae 
becomes important.

%***************************************************************************
\item Furthermore, it has been demonstrated that
supernova explosions create a cloudy medium by enhancing the
density and thus the cooling in the shock shell, as well as providing
a hot, low-density gas.  Therefore, we have provided
a simplified description of our results,
which allows a more realistic treatment of energy input and the
mass redistribution
by supernovae.

\end{enumerate}

%***************************************************************************

This research began as an attempt to provide an improved measure of the 
consequences of supernova energy input for the class of models of galactic 
evolution which attempt to take heating and cooling effects into 
account properly. There are indeed a number of important problems for which a 
careful treatment of supernova input is essential. This includes the 
formation and early evolution of galaxies, where  
supernova energy input may cause significant mass loss, 
as possibly reflected in the approximately solar metallicity of the hot gas 
observed in X-ray emission from clusters of galaxies (Mushotzky 
{\it et~al.}\markcite{Mush1996} 1996). 
In addition, the supernova energy input is likely to have implications for 
the abundance evolution of starburst 
nucleus galaxies (Coziol\markcite{Cozi1996} 1997). In the context of models of 
self-enrichment, it may also prove to be relevant to the interpretation of 
the metallicity distributions of globular clusters.  Furthermore, energy input
due to supernovae has been suggested to cure some shortcomings in hierarchical
scenarios of galaxy formation, e.g. the overcooling problem (White \& Rees 1978)
or the angular momentum problem of disk galaxies (Navarro \& Steinmetz 1997). 

A critical implication of our results  
is the quantity of metals produced per unit energy input by supernovae; 
if the realistic energy input 
from a supernova is $\approx~10^{50}$ ergs rather than $\approx~10^{51}$ ergs, 
approximately 10 times as much metals may be produced per 
unit supernova energy input.  If supernovae are the major source of energy,
this result has direct implications for the interplay of chemical and dynamical
evolution of the environment.

%***************************************************************************

\begin{acknowledgements}
We are grateful to James Truran for the support throughout the development 
and the completion of this project.  We thank Angela Olinto and Louis
Tao for their comments on the manuscript.  Tomasz Plewa also provided
helpful comments and discussion.  K.~T. is grateful to the National Science
Foundation for support through grant AST 92-17969.  
H.-Th.~J. was supported in part
by the National Science Foundation under grant
NSF AST 92-17969, by the National Aeronautics and Space
Administration under grant NASA NAG 5-2081, and by an
Otto Hahn Postdoctoral Scholarship of the Max-Planck-Society.
This work was also partially supported by the
Sonderforschungsbereich SFB 375-95 ``Astro-Teilchenphysik'' der
Deutschen
Forschungsgemeinschaft.

\end{acknowledgements}

\end{document}